\documentclass[draft]{agujournal2019}
\usepackage{url} 
\usepackage{lineno}
\usepackage[finalnew]{trackchanges} 
\usepackage{soul}
\linenumbers
\usepackage{amsmath}
\usepackage{dsfont}
\usepackage{tikz}

\drafttrue

\journalname{Space Physics}

\begin{document}
\nolinenumbers

\newcommand{\divB}{\mbox{$\boldsymbol{\nabla} \boldsymbol{\cdot} \mathbf{B}$}} 

\newcommand\B{\mathbf{B}}
\newcommand\J{\mathbf{j}}
\newcommand\x{\mathbf{x}}
\newcommand\BS{\text{BS}}
\newcommand\HH{\text{H}}
\newcommand\M{\mathcal{M}}
\newcommand\MI{\mathcal{M}_{\text{I}}}
\newcommand\MO{\mathcal{M}_{\text{O}}}
\newcommand\G{\mathcal{G}}
\newcommand\I{\partial{\mathcal{G}}}
\newcommand\n{\mathbf{\hat{n}}}

\newcommand\diverr{\mathbf{B}_{\text{div}}}
\newcommand\outerr{\mathbf{B}_{\text{out}}}
\newcommand\innerint{\mathbf{B}_{\text{in}}}
\newcommand\bsint{\mathbf{B}_{\text{BS}}}

\newcommand\LCdiverr{{B}_{\text{div}}}
\newcommand\LCouterr{{B}_{\text{out}}}
\newcommand\LCinnerint{{B}_{\text{in}}}
\newcommand\LCbsint{{B}_{\text{BS}}}

\newcommand\ifpi{\frac{1}{4\pi}}

\newcommand\dive[1]{\boldsymbol{\nabla} \boldsymbol{\cdot} #1}
\newcommand\curl[1]{\boldsymbol{\nabla} \times #1}

\newcommand\normal[1]{\hat{\mathbf{n}}_{#1}}

\newcommand\dV{\mathrm{d}^3x}

\newcommand\xoverx{\frac{\phantom{|} \x_0-\x \phantom{|^3}}{| \x_0-\x |^3}}

\newcommand\surf[2]{\bigg(\xoverx [#2 \boldsymbol{\cdot} #1] + \xoverx\times [#2 \times #1] \bigg)}

\newcommand\coulombInt[1]{\ifpi \int_{#1} [\dive{\B}(\x)] \xoverx \dV}
\newcommand\coulombIntTilde[1]{\ifpi \int_{#1} \dive{\tilde{\B}}(\x) \xoverx \dV}

\newcommand\biotsavInt[1]{\ifpi \int_{#1} \big[ \curl{\B}(\x) \big] \times \xoverx \dV}
\newcommand\biotsavIntTilde[1]{\ifpi \int_{#1} \big[ \curl{\tilde{\B}}(\x) \big] \times \xoverx \dV}
\newcommand\biotsavIntJ[1]{\frac{\mu_o}{4\pi} \int_{#1} \J_{\text{FAC}}(\x) \times \xoverx \dV}

\newcommand\surfInt[1]     {\ifpi \oint_{#1} \surf{\normal{#1}}{\B(\x)} dS}
\newcommand\surfIntNEG[1]  {\ifpi \oint_{#1} \surf{(-\normal{#1})}{\B(\x)} dS}
\newcommand\surfIntTilde[1]{\ifpi \oint_{#1} \surf{\normal{#1}}{\tilde{\B}(\x)} dS}

%
%


\title{$\dive\mathbf{B}$, outer boundary, and Biot--Savart in magnetosphere MHD simulations}

%
%




\authors{Dean Thomas\affil{1}, Robert S. Weigel\affil{1}, Gary Quaresima\affil{2}, Antti Pulkkinen\affil{3}, Daniel T. Welling\affil{4}}

\affiliation{1}{Space Weather Lab, Department of Physics and Astronomy, George Mason University, Fairfax, VA, USA}
\affiliation{2}{Department of Physics, University of Virginia, Charlottesville, VA, USA}
\affiliation{3}{Heliophysics Science Division, NASA Goddard Space Flight Center, Greenbelt, MD, USA}
\affiliation{4}{Climate \& Space Research, University of Michigan, Ann Arbor, MI, USA}




\correspondingauthor{Dean Thomas}{dthoma6@gmu.edu}



\begin{keypoints}
\item $\dive\mathbf{B} \ne 0$ and outer surface boundary integrals affect Biot-Savart analysis of magnetospheric magnetohydrodynamic results.
\item These integrals are frequently ignored when computing the magnetospheric contribution to the magnetic field on Earth's surface.
\item\change[DT]{ However, under some circumstances, they make significant contributions to magnetic field calculations.}{ Instead of Biot-Savart, the integral over the inner surface boundary should be used.}
\end{keypoints}

%
%


\begin{abstract}
We examine the size of {\divB} and outer surface boundary integrals in estimating the surface magnetic field from magnetohydrodynamic (MHD) simulations. Maxwell’s equations tell us \mbox{$\dive\mathbf{B} = 0$}, which may be violated due to numerical error. MHD models such as the Space Weather Modeling Framework (SWMF) and the Open Geospace General Circulation Model (OpenGGCM) use different techniques to limit {\divB}. Analyses of MHD simulations typically assume {\divB} errors are small. Similarly, analyses commonly use the Biot--Savart Law and magnetospheric current density estimates from MHD simulations to determine the magnetic field at a specific point on Earth. This calculation frequently omits the surface integral over the outer boundary of the simulation volume that the Helmholtz decomposition theorem requires. This paper uses SWMF and OpenGGCM simulations to estimate the magnitudes of the {\divB} and outer boundary integrals compared to Biot--Savart estimates of the magnetic field on Earth. In the simulations considered, the {\divB} and outer surface integrals are up to $30$\% of Biot--Savart estimates when the Biot--Savart estimates are large.  We conclude rather than using the Biot--Savart Law to estimate the magnetic field from the magnetosphere, it is better and computationally more efficient to use the integral over the inner boundary of the magnetosphere. The conclusions are the same for a simulation involving a simple change in the interplanetary magnetic field and a more complex superstorm simulation. 
\end{abstract}

%
%

\section{Introduction}

In this paper, we examine two integrals -- the {\divB} volume integral and the surface integral over the outer boundary of the simulation volume -- that affect estimates of the magnetic field on \remove[DT]{the }Earth's surface using magnetohydrodynamic (MHD) simulation results. 

According to Maxwell's equations, \mbox{$\dive\mathbf{B} = 0$}; however, this condition may be violated due to numerical error in MHD simulations \cite{Toth2000}. To combat this issue, MHD models use numerical techniques to limit {\divB}. Consequently, researchers typically ignore {\divB} errors \cite{rastatter2014}. 

It is also common to apply the Biot--Savart Law to MHD magnetospheric current density estimates, $\mathbf{j}$, to determine the magnetic field at a specific point, such as a magnetometer site on \remove[DT]{the }Earth’s surface \cite{rastatter2014}. Consistency with the Helmholtz decomposition theorem requires that the surface integral over the outer boundary of the simulation volume be included, but it is frequently omitted in these calculations.  

There are other concerns with \change[DT]{applying the Biot--Savart Law}{analysis of MHD results}.  \citeA{parker1996} notes that this approach is ineffective when treating strong, non-symmetric, or time-dependent situations.  And \citeA{ferdousi2016} saw substantial signal propagation times in MHD results (e.g., 72 seconds for propagation over 30 Earth radii from the tail to Earth).  All of these factors affect the analysis of MHD results but are beyond the scope of this paper.

This paper examines {\divB} and the outer boundary integrals using two MHD models to estimate their magnitude compared to other magnetospheric contributions to the magnetic field. There are differences between the two models examined due to differing methods of handling {\divB}, solving the MHD equations, and modeling the ionosphere. While the two models provide similar results, they are not identical.  

The integrals depend on the size of the Biot--Savart estimates of magnetospheric contributions to the magnetic field on Earth.  In the simulations considered, the {\divB} and outer surface integrals are up to $30$\% of Biot--Savart estimates when the Biot--Savart estimates are large.  The percentages increase when the Biot--Savart estimates are small.  The contributions from the outer surface boundary term are typically larger than contributions from the {\divB} term.  The integrals are spatially correlated -- locations on \remove[DT]{the }Earth's surface near one another exhibit contributions from these integrals of similar magnitudes. The conclusions are the same for a simple simulation involving a change in the interplanetary magnetic field and a more complex simulation representing a superstorm. 

\add[DT]{As we will show below, concerns with $\divB$ and the outer boundary integrals can be avoided.  It is easier and more computationally efficient to calculate the magnetic field due to the magnetosphere using the integral over the inner boundary of the magnetosphere rather than applying the Biot--Savart Law to the magnetospheric current densities.  Using the inner boundary integral avoids concerns with $\divB$ and the outer boundary integrals.}

\section{Background}

\begin{figure}
    \begin{center}
%
           \includegraphics[width=0.8\textwidth]{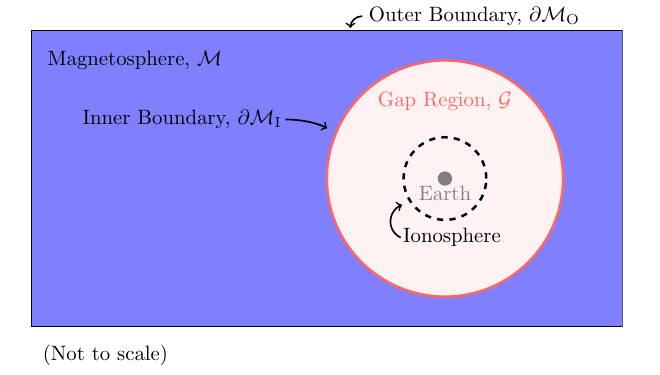}
    \end{center}
    \caption{Schematic representation of MHD simulation volume. The blue rectangle is the magnetosphere. The pale red circle encloses the gap region. The dashed black circle is the ionosphere. The small gray circle is Earth. The outer and inner magnetosphere boundaries are identified.}
    \label{fig:schematic}
\end{figure}

To set the stage for our discussion, we refer to Figure~\ref{fig:schematic}, which provides a schematic representation of an MHD--based magnetosphere model. We examine a simulation of a region around Earth. The simulation encompasses a large volume, $\mathcal{M}$, representing the magnetosphere. Inside it, there is a smaller gap region, $\mathcal{G}$, where a dipole magnetic field and field--aligned currents are assumed to reduce simulation time \cite{Yu_2010}. This region contains ionospheric currents on a spherical shell centered on Earth and continues down to \remove[DT]{the }Earth's surface. Earth is at the center of the gap region. In our notation, $\partial \mathcal{M}$ is the boundary of $\mathcal{M}$ and includes both the outer ($\partial \mathcal{M}_{O}$) and inner ($\partial \mathcal{M}_{I}$) boundaries. $\partial \mathcal{G}$ is the outer boundary of $\mathcal{G}$.  It coincides with $\partial \MI$ and has a surface normal in the opposite direction, $\normal{\partial \G} = - \normal{\partial \MI}$. 

To understand the relationship between the Biot--Savart Law and the {\divB} and outer boundary integrals, we employ the Helmholtz decomposition theorem for a vector field \cite{Arfken2005}. We examine two cases, applying the theorem to volumes $\mathcal{M}$ and $\mathcal{G}$ to determine the magnetospheric contributions\add[DT]{ in excess of Earth's intrinsic dipole field} to the magnetic field inside $\mathcal{G}$, such as a point on Earth's surface. 

\subsection{Helmholtz Decomposition Theorem Applied to $\mathcal{M}$}
\label{section:Helmholtz Magnetosphere}

The Helmholtz decomposition theorem states that a vector field, in this case the magnetic field $\mathbf{B}$, can be written as the sum of irrotational and solenoidal components. For $\mathbf{B}$ defined over a finite--volume, $\mathcal{M}$, and with \mbox{$\x_0 \in \mathcal{M}$}, the theorem states that for sufficiently smooth $\mathbf{B}$:
\linebreak
\begin{equation}
\begin{split}
\mathbf{B}(\mathbf{x}_0)
  &= \phantom{+}   \coulombInt{\M} \\
  &  \phantom{=} + \biotsavInt{\M}  \\
  &  \phantom{=} - \surfInt{\partial \M} 
\end{split}
\label{eqn:hdt}
\end{equation}
\linebreak
This equation involves volume integrals over $\mathcal{M}$ and surface integrals over the boundary $\partial \mathcal{M}$. The $\curl \mathbf{B}$ volume integral is the Biot--Savart Law contribution: 
\begin{equation}
 \B_{\BS}(\x_0) := \biotsavInt{\M}
\label{eqn:biotsimple}
\end{equation}

A \change[DT]{complimentary}{complementary} result is valid for $\mathbf{x}_0 \not \in \mathcal{M}$, for example, $\mathbf{x}_0 \in \mathcal{G}$ \cite{Gombosi2021}.  In this case, the left--hand side of equation~\ref{eqn:hdt} is 0.  One way to derive the Helmholtz decomposition theorem starts with the delta function: \mbox{$\mathbf{B}(\mathbf{x}_0) = \int_{\mathcal{M}} \mathbf{B}(\mathbf{x}) \delta(\mathbf{x} - \mathbf{x}_0) d^{3}x$}. From this, it follows that $\B(\x_{0})$ is 0 for $\mathbf{x}_0 \not \in \mathcal{M}$. 

With the left--hand side 0, equation~\ref{eqn:hdt} can be rearranged to isolate the $\bsint$ term. This version is useful for examining the Biot--Savart Law applied at a point inside $\mathcal{G}$, such as a magnetometer site on \remove[DT]{the }Earth's surface:
\linebreak
\begin{equation}
\begin{split}
\label{eqn:biot M}
\B_{\BS}(\x_0)
  &= - \coulombInt{\M} \\
  &  \phantom{=} + \surfInt{\partial \M} 
\end{split}
\end{equation}
\linebreak
$\partial \mathcal{M}$ can be split into outer and inner boundaries with $\partial \M = \partial \MO + \partial \MI$. $\partial \G$ coincides with $\partial \MI$ with the opposite normal direction. Thus, equation \ref{eqn:biot M} can be written as:
\begin{equation}
\begin{split}
\label{eqn:biot M expanded} 
\B_{\BS}(\x_0)
  &=  - \coulombInt{\M} \\
  &  \phantom{=} + \surfInt{\partial \MO} \\
  &  \phantom{=} - \surfInt{\partial \G}
\end{split}
\end{equation}
This equation is equivalent to equation~\ref{eqn:biotsimple}.

\subsection{Helmholtz Decomposition Theorem Applied to $\mathcal{G}$}
\label{section:Helmholtz Gap}

We arrive at a different equation when the Helmholtz decomposition theorem is applied to volume $\mathcal{G}$ with $\mathbf{x}_0 \in \mathcal{G}$. Inside of $\mathcal{G}$, $\mathbf{B}$ approximates a dipole field, with field--aligned and ionospheric currents in $\mathcal{G}$ generating deviations from a dipole. Thus, in MHD simulations, inside $\G$, $\mathbf{B}$ is commonly treated as a dipole magnetic field with $\dive{\mathbf{B}} = 0$ and $\curl \mathbf{B} = 0$ \cite{Yu_2010}.  The deviations due to field--aligned and ionospheric currents typically are handled by applying the Biot-Savart Law to these currents \cite{rastatter2014}.  Ignoring these deviations for the moment, for {$\x_0 \in \G$}, the Helmholtz decomposition theorem gives:   
\begin{equation}
\begin{split}
\label{eqn:biot G}
\innerint(\x_0)
  &= - \surfInt{\partial \G}
\end{split}
\end{equation}
where $\innerint$ is the integral over the inner surface of the magnetosphere, and along with equation~\ref{eqn:biot M expanded}, is a second way of computing the magnetospheric current contribution to $\B$ on Earth. 

\add[DT]{Using $\innerint$ to determine magnetospheric contributions is more efficient and easier to calculate.  Numerically, it is faster to calculate the surface integral over the inner boundary rather than integrating across the magnetospheric volume.  And the result is independent of $\divB$ and the outer boundary of the magnetosphere.}

\subsection{Deviations in Gap Region}
\label{section:Deviations in Gap Region}
The deviations to the dipole field described in section~\ref{section:Helmholtz Gap} typically are handled by applying the Biot-Savart Law to the field--aligned and ionospheric currents in $\G$.  The field--aligned current density, $\J_{\text{FAC}}$, creates a deviation to $\B(\x_0)$ of the form:
\begin{equation*}
 \B_{\text{FAC}}(\x_0) := \biotsavIntJ{\G}
\end{equation*}
There is a similar equation, $\B_{\text{iono}}$, due to ionospheric currents that also are handled via the Biot--Savart Law \cite{rastatter2014}.

$\B_{\text{FAC}}$ and $\B_{\text{iono}}$ are divergence-free and do not introduce {\divB} integrals.  This fact is easily demonstrated.  For \mbox{$\dive\B_{\text{FAC}}$}, the divergence derivatives are over $\x_0$, and the Leibniz rule allows the divergence to be moved inside the integral.  We then apply the vector identity:
\begin{equation*}
\dive{(\mathbf{A} \times \mathbf{B})} = (\curl \mathbf{A}) \cdot \mathbf{B} - \mathbf{A} \cdot (\curl \mathbf{B})
\end{equation*}
to the integrand. \mbox{$\curl\J_{\text{FAC}} = 0$} because the divergence derivatives are over $\x_0$ and $\J_{\text{FAC}}$ only depends on $\x$.  Expanding the derivatives shows
\begin{equation*}
\curl\frac{\x - \x_0}{|\x - \x_0|^3} = 0
\end{equation*}  With both terms vanishing, \mbox{$\dive{ \B_{\text{FAC}}} = 0$}.  The same argument demonstrates \mbox{$\dive\B_{\text{iono}} = 0$}.

 To determine the total change in $\B$ on Earth due to magnetospheric, field--aligned, and ionospheric currents, the deviations are added to either equation~\ref{eqn:biot M expanded} or \ref{eqn:biot G}.  The $\B_{\text{FAC}}$ and $\B_{\text{iono}}$ deviations are identical for either equation.

\subsection{Comparing Equations}
\label{section:comparing}

The relationship between the {\divB} and outer surface boundary integrals and the application of the Biot--Savart Law to magnetospheric currents is understood if we compare equations~\ref{eqn:biot M expanded} and \ref{eqn:biot G}. We begin by defining divergence and outer boundary integrals:
\begin{equation}
\label{eqn:intdiv}
\diverr(\x_0) :=  - \coulombInt{\M}
\end{equation}
and
\begin{equation}
\label{eqn:intout}
\outerr(\x_0) :=  + \surfInt{\partial \MO}
\end{equation}
With these definitions, the difference between equations~\ref{eqn:biot M expanded} and \ref{eqn:biot G} is:
\begin{equation}
\label{eqn:difference}
  \bsint(\x_0) - \innerint(\x_0) = \diverr(\x_0) +  \outerr(\x_0)
\end{equation}
We see that $\bsint(\x_0)$ reduces to $\innerint(\x_0)$ when $\diverr(\x_0)$ and $\outerr(\x_0)$ are small. 
\citeA{Gombosi2021} derived a similar equation (C.43 in their paper)\add[DT]{ that describes the SWMF implementation of $\innerint$}.  However, their derivation assumes the outer boundary is at infinity, the magnetic field decays sufficiently rapidly at large distances, and {\divB} is zero. Thus, their equation only includes $\innerint$ with  $\outerr=0$ and $\diverr=0$.

\add[DT]{We note that $\outerr$ is the contribution from the currents outside the MHD domain that are ignored in the traditional Biot--Savart approach. Since $\outerr$ is an integral over the outer boundary of the simulation domain, it is affected by the size of the domain for a given scenario.  Differing domains may change the results and $\outerr$.}

\add[DT]{A similar statement is true for $\innerint$.  While in the simulations that we examine, the inner boundary of the magnetosphere is the same, this is not necessarily true.  Model users could specify different inner boundaries.  As with $\outerr$, differing simulation domains may affect simulation results and the integrals based on those results.}

\section{Models and Solar Wind Conditions}

To estimate the magnitude of the $\diverr$ and $\outerr$ integrals, we use two \change[DT]{MHD}{global magnetosphere} models for the analysis: Space Weather Modeling Framework (SWMF) \cite{Gombosi2003adaptive} and Open Geospace General Circulation Model (OpenGGCM) \cite{Raeder2008openggcm}. They were chosen because they use different approaches to solve the MHD equations and to control {\divB} numerical errors. 

For SWMF, the Block--Adaptive--Tree--Solarwind--Roe--Upwind--Scheme (BATS--R--US) model is used to simulate the magnetosphere.  It is a generalized MHD code that utilizes adaptive mesh refinement on a three--dimensional Cartesian grid. To solve the MHD equations, it employs \change[DT]{Roe's}{an} approximate Riemann solver. For controlling {\divB} terms, BATS--R--US uses the eight--wave scheme based on a symmetric form of the MHD equations. \change[DT]{While computationally inexpensive to implement, the eight-wave scheme has a known limitation, $\diverr$ may grow in regions where the source is large}{The eight-wave scheme is computationally inexpensive to implement. It calculates $\diverr$ at the truncation level, but $\diverr$ may be significant near discontinuities, such as shocks} \cite{Gombosi2003adaptive, Powell1999solution}.

OpenGGCM is a global magnetosphere model that uses a semi--conservative form of the MHD equations on a three-dimensional, stretched Cartesian grid. The equations conserve plasma energy and use a second-order predictor--corrector finite difference scheme for solving the MHD equations. To control {\divB} terms, OpenGGCM uses the Constrained Transport (CT) method that employs a staggered grid that maintains \mbox{$\dive\mathbf{B} = 0$} to round--off error. Historically, the primary disadvantage of CT is that it is not easy to extend the method to generalized grids \cite{Evans1988simulation}. However, there has been work on this limitation \cite{mignone2021}.

We use the same solar wind conditions in both models. Because the integrals in equations~\ref{eqn:biot M expanded} and \ref{eqn:biot G} depend on the magnetic field in the magnetosphere, we use solar wind conditions that produce significant changes to it. Consequently, all solar wind conditions except the $Z$--component of the interplanetary magnetic field, $B_Z^\text{\tiny{IMF}}$, are held constant and near their long--term average values \cite{Curtis2014}.  The simulation runs from 00:00 to 20:00 UTC, and the dipole tilt is zero and constant.  Ion number density, $n$, is 5$/\text{cm}^{3}$. Temperature, $T$, is $10^{5}\text{ K}$. The $X$--component of solar wind velocity, $V_X$, is $-400\text{ km/s}$. $B_Z^\text{\tiny{IMF}}$ is $+5\text{ nT}$ from 00:00 until 06:00. At 06:00, $B_Z^\text{\tiny{IMF}}$ flips to $-10\text{ nT}$ and remains at this value until 20:00. All other parameters ($V_Y$, $V_Z$, $B_X^\text{\tiny{IMF}}$, and $B_Y^\text{\tiny{IMF}}$) are 0. All solar wind conditions are in Geocentric Solar Magnetospheric (GSM) coordinates at 33 Earth radii ($R_E$) sunward of Earth.

The model runs were executed at NASA's Community Coordinated Modeling Center (CCMC; \citeA{Hesse2001}).

\section{Magnitude of integrals}
\label{section:Magnitude term}

Using equations \ref{eqn:biotsimple} through \ref{eqn:difference} and results from the MHD simulations, we calculate $\mathbf{B}$ at points on \remove[DT]{the }Earth's surface. We determine the magnetospheric contributions from the four integrals in these equations: $\bsint$, $\innerint$, $\outerr$, and $\diverr$. For $\bsint$ (Eqn.~\ref{eqn:biotsimple}), we use the $\mathbf{j}$ provided in the MHD simulation output files to replace \mbox{$\curl \mathbf{B}/\mu_0$}.  (We verified that $\mathbf{j}$ = $\curl \mathbf{B}/\mu_0$.) For the surface integrals, $\innerint$  (Eqn.~\ref{eqn:biot G}) and $\outerr$ (Eqn.~\ref{eqn:intout}), we interpolate the integrands over the inner and outer boundaries of the magnetosphere volume. $\partial \MI$ is a sphere of radius $3 R_E$ and $\partial \MO$ is a rectangular \change[DT]{prism}{domain} for both simulations:  
\begin{itemize}
    \item The BATS--R--US \change[DT]{rectangular prism}{domain} is $-224$~$R_E$ to 32~$R_E$ in the $X$--direction and $-128$~ $R_E$ to 128~$R_E$ in the $Y$-- and $Z$--directions (GSM).
    \item The OpenGGCM \change[DT]{rectangular prism}{domain} is $-350$~$R_E$ to 60~$R_E$ in the $X$--direction and $-48$~ $R_E$ to 48~$R_E$ in the $Y$-- and $Z$--directions (GSE).
\end{itemize}
For $\diverr$ (Eqn.~\ref{eqn:intdiv}), we use second--order stencils \cite{singh2009finite} to calculate {\divB}. 

\begin{figure}[!htb]
\includegraphics[width=\textwidth]{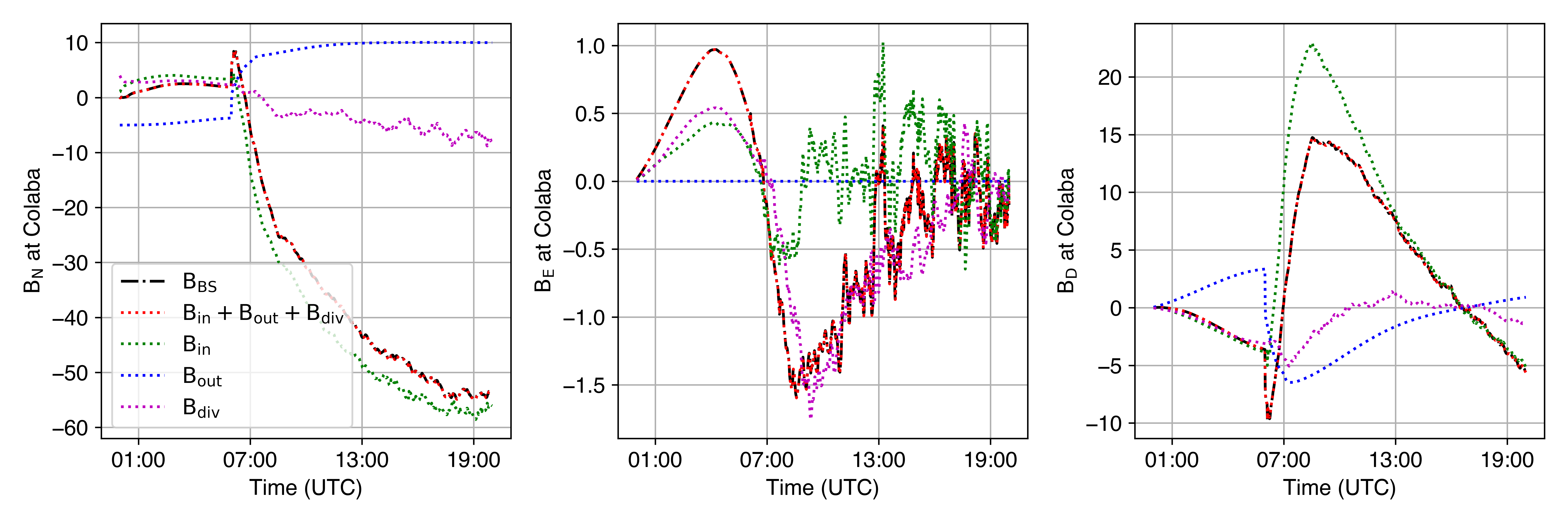}
\caption{$\bsint$, $\innerint$, $\diverr$, and $\outerr$ versus time derived from BATS--R--US simulation. Plots show $\mathbf{B}$ at Colaba, India. $\innerint$ (blue line) explains most of the variation in $\bsint$; with $\bsint$ (black line) equal to the sum of $\innerint$, $\outerr$, and $\diverr$ (red line).}
\label{fig:BATSRUS point}
\vspace{5mm}
\includegraphics[width=\textwidth]{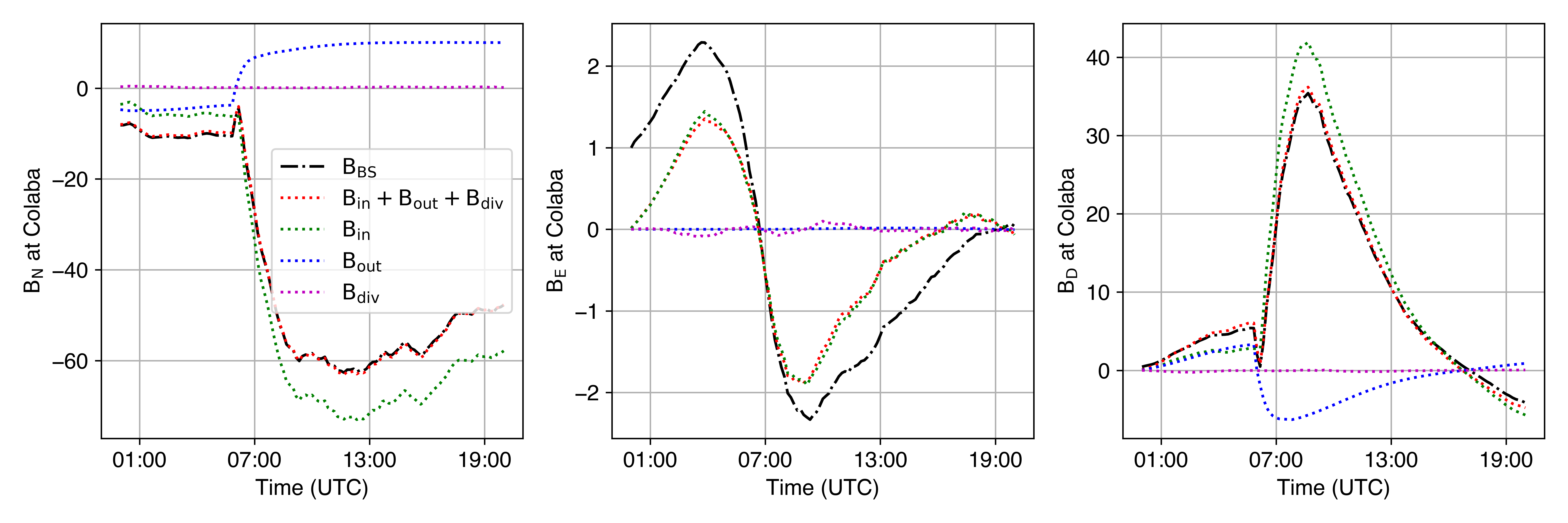}
\caption{$\bsint$, $\innerint$, $\diverr$, and $\outerr$ versus time derived from OpenGGCM simulation. Plots show $\mathbf{B}$ at Colaba, India. As in Figure~\ref{fig:BATSRUS point}, $\innerint$ (blue line) explains most of the variation in the $\bsint$; with $\bsint$ (black line) approximately equal to the sum of $\innerint$, $\outerr$, and $\diverr$ (red line).  The difference between the black and red lines is due MHD discontinuities, see section~\ref{section:Discontinuities}.}
\label{fig:OpenGGCM point}
\end{figure}

In examining the results, we use a multi--prong approach. At a mid-latitude magnetometer site, we calculate the $\mathbf{B}$ contributions from the four integrals discussed above. Knowing the relationship defined by equation~\ref{eqn:difference}, we look at the contributions from the integrals to identify the overall characteristics and the trends in the results. Next, we extend this analysis across \remove[DT]{the Earth's surface to determine whether the trends and relationships observed at a single magnetometer site are similar across the }Earth. As part of this analysis, we examine the relative size of the integrals. Finally, we compare the magnitudes of $\innerint$, $\diverr$, and $\outerr$ to $\bsint$ to determine additional correlations in the data.

We select Colaba, India, as the mid--latitude magnetometer site.  It was chosen simply because we have examined this magnetometer site in other analyses \cite{Thomas2024}. $\mathbf{B}$ at Colaba, broken down into \remove[DT]{various combinations of} $\bsint$, $\innerint$, $\diverr$, and $\outerr$, is shown in Figures~\ref{fig:BATSRUS point} and \ref{fig:OpenGGCM point} for the two simulations. Before the $B_Z^\text{\tiny{IMF}}$ flip at 06:00, $\mathbf{B}$ is small in both simulations. After the $B_Z^\text{\tiny{IMF}}$ flip, both simulations show significant changes in $\mathbf{B}$.

Overall, BATS--R--US and OpenGGCM provide \change[DT]{similar, but not identical,}{qualitatively similar} results. The curves have similar shapes, but different magnitudes. The largest differences are in \add[DT]{$B_E$ and }$B_D$, which differ by up to a factor of $\approx2$. Although BATS--R--US and OpenGGCM are both MHD--based models, as discussed in section~\ref{section:comparing}, their approaches differ.  Among other differences, their approaches to solving the MHD equations and their ionosphere models differ.  So differences in their predictions of  $\B$ are expected. 

With respect to equation~\ref{eqn:difference}, which tells us that $\bsint$ reduces to $\innerint$ when $\diverr$ and $\outerr$ are small, we make similar observations for the two simulations. For both, $\innerint$ explains most of the variation in $\bsint$. For BATS--R--US, we find that equation~\ref{eqn:difference} is satisfied (as demonstrated by the overlapping black and red lines in Figure~\ref{fig:BATSRUS point}). However, for OpenGGCM we see a small difference (the black and red lines are close, but not always overlapping in Figure~\ref{fig:OpenGGCM point}). \remove[DT]{This apparent discrepancy is due to coordinate transformations and decomposition of $\mathbf{B}$ into North, East, and Down components. BATS--R--US uses the Geocentric Solar Magnetospheric (GSM) coordinate system, whereas OpenGGCM uses the Geocentric Solar Ecliptic (GSE) coordinate system.  To facilitate comparison between the model results, we transformed the OpenGGCM results into GSM.  The coordinate system transformation and vector decomposition cause the discrepancies between the solid black and red lines in Figure~\ref{fig:OpenGGCM point}. This is demonstrated in Figure~\ref{fig:OpenGGCM mag point}, which shows OpenGGCM $|\mathbf{B}|$.  In this figure, the black and red lines overlap, consistent with equation~\ref{eqn:difference}.}


\section{Effect of Non-Smooth Fields}
\label{section:Discontinuities}

\add[DT]{The small difference between the black and red lines in }Figure~\ref{fig:OpenGGCM point}\add[DT]{ is due to MHD discontinuities in OpenGGCM magnetic field. The difference is most apparent for $B_E$, for which it is $\approx 1$ (nT).} 

\add[DT]{The Helmholtz decomposition theorem requires that the vector field be sufficiently smooth. However, the OpenGGCM magnetic field has spikes not seen the BATS--R--US results. }Figure~\ref{fig:Compare B fields} \add[DT]{ shows the $x$, $y$, and $z$ components (GSM) of the BATS--R--US and OpenGGCM magnetic fields measured along a line parallel to the $x$ axis. The OpenGGCM field has narrow spikes. These spikes are due to MHD shocks in the simulation, and represent a discontinuity that violates the sufficiently smooth vector field requirement for the Helmholtz decomposition theorem.}  

\begin{figure}[!htb]
\begin{center}
    \includegraphics[width=\textwidth]{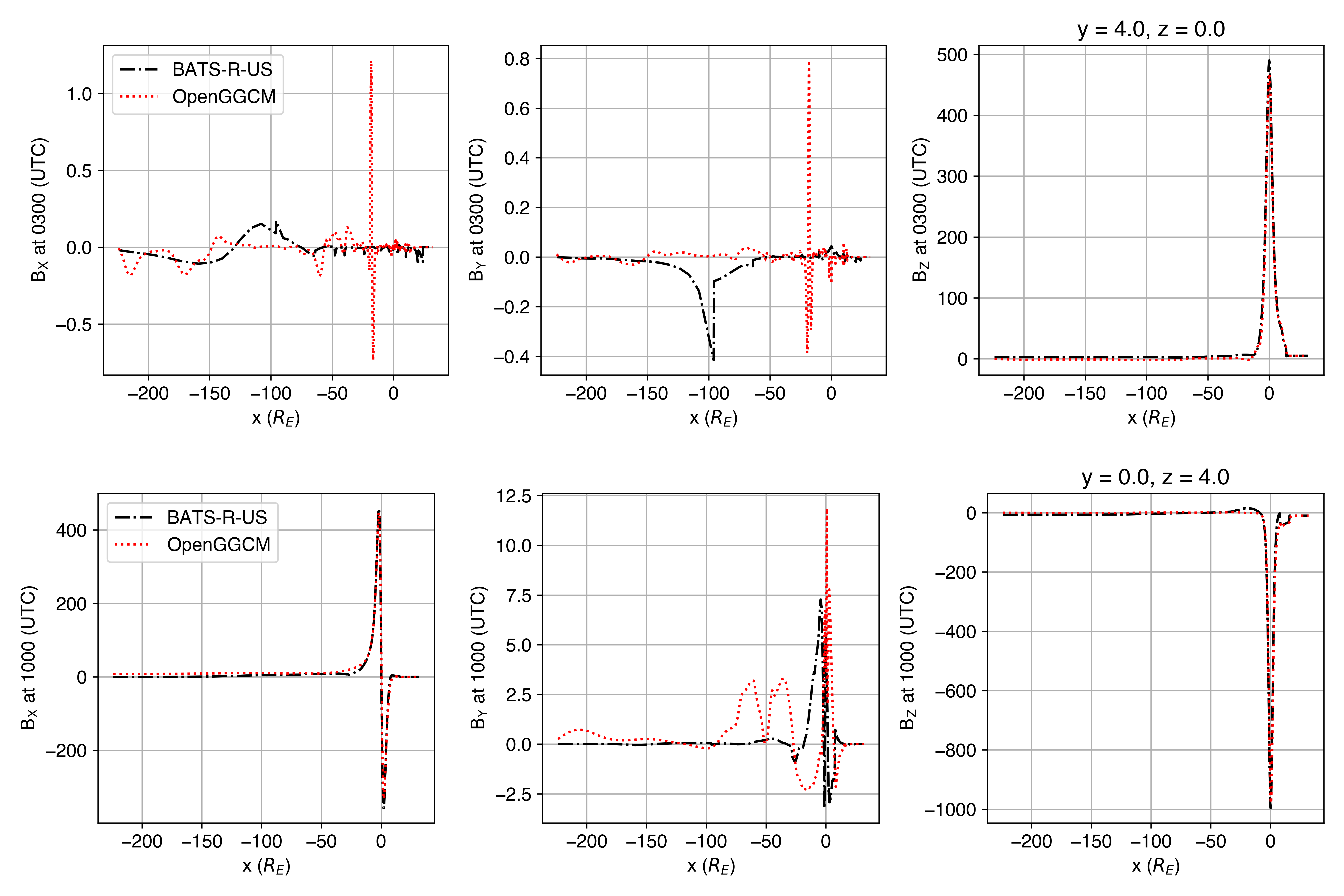}
\end{center}
\caption{The $x$, $y$, and $z$ components (GSM) of the BATS--R--US and OpenGGCM $\mathbf{B}$ measured along a line parallel to the $x$ axis. $y$ and $z$ values shown in titles.  The top graph is at 03:00 (UTC), the bottom at 10:00 (UTC).   OpenGGCM has narrow spikes near $x=-20$ in the top graph and near $x=0$ in the bottom graph.}
\label{fig:Compare B fields}
\end{figure}

\add[DT]{Two test cases used to validate our software highlight the effect of discontinuities. The first test case involves a smooth field without discontinuities.  The test involves an infinite line current in the $x-y$ plane, running parallel to the $x$ axis at $y=-256R_E$. The magnetic field and associated current density are placed on the BATS--R--US and OpenGGCM grids, }Figures~\ref{fig:BATSRUS line current} and ~\ref{fig:OpenGGCM line current}.\add[DT]{  Since $\nabla \cdot \mathbf{B} = 0$ and $\nabla \times \mathbf{B} = 0$ inside the simulation domains, we expect $\bsint$ and $\diverr$ to be 0.  Furthermore, the Helmholtz decomposition theorem, }equation~\ref{eqn:difference},\add[DT]{ gives the result that $\innerint = -\outerr$ under these conditions.  The data in }figures~\ref{fig:BATSRUS line current} and ~\ref{fig:OpenGGCM line current} \add[DT]{are consistent with these expectations. We see that the Helmholtz decomposition theorem holds, the black and red lines overlap.}

\add[DT]{The second test case involves a discontinuity, a $B_Z^\text{\tiny{IMF}}$ flip.  $B_Z^\text{\tiny{IMF}}$ is $-$15 (nT) for $y>\pi$, and 5 (nT) otherwise. $\pi$ was selected so that the discontinuity occurs between grid points. As with the previous test case, the magnetic field and current density are placed on the BATS--R--US and OpenGGCM grids, }Figures~\ref{fig:BATSRUS Bz flip} and \ref{fig:OpenGGCM Bz flip}.\add[DT]{  The magnetic field is discontinuous at $y=\pi$, thus the Helmholtz Decomposition Theorem does not be apply. These figures demonstrate this fact. The black and red lines do not overlap.}

\begin{figure}[!htb]
\begin{center}
    \includegraphics[width=\textwidth]{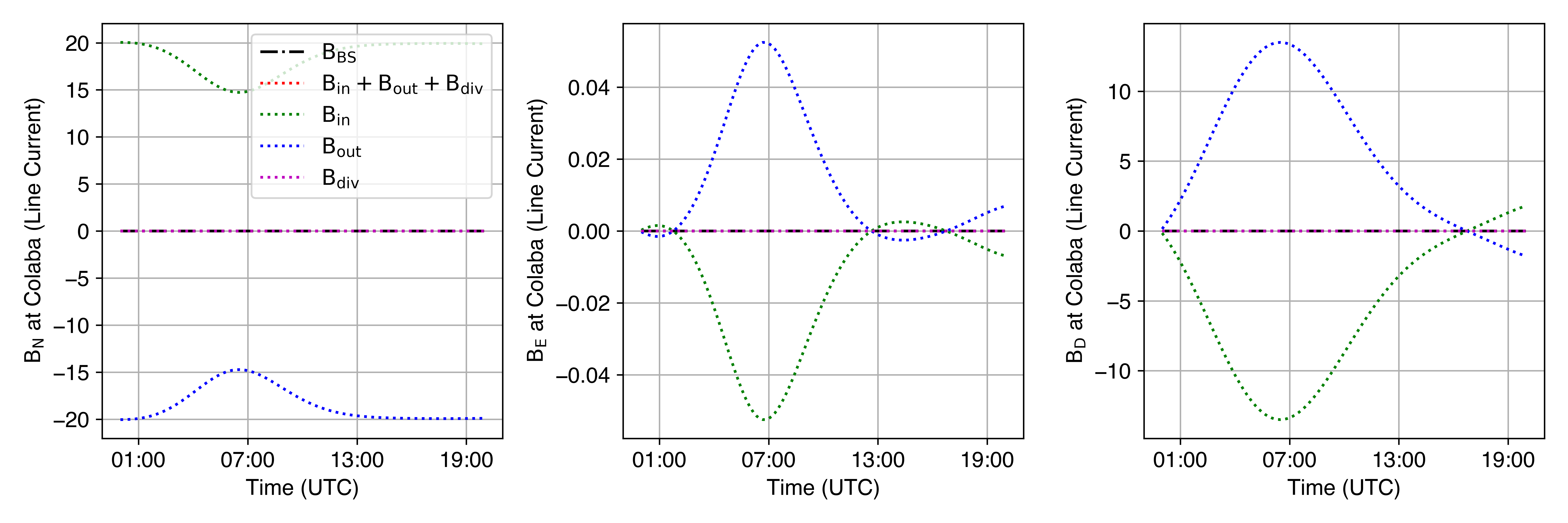}
\end{center}
\caption{$\bsint$, $\innerint$, $\diverr$, and $\outerr$ versus time derived from BATS--R--US simulation grid assuming infinite line current outside the simulation domain. Plots show $\mathbf{B}$ at Colaba, India. Since $\nabla \cdot \mathbf{B} = 0$ and $\nabla \times \mathbf{B} = 0$, we expect $\bsint = 0$ and $\diverr = 0$.  In addition, the Helmholtz decomposition theorem requires that $\innerint = - \outerr$.}
\label{fig:BATSRUS line current}
\begin{center}
    \includegraphics[width=\textwidth]{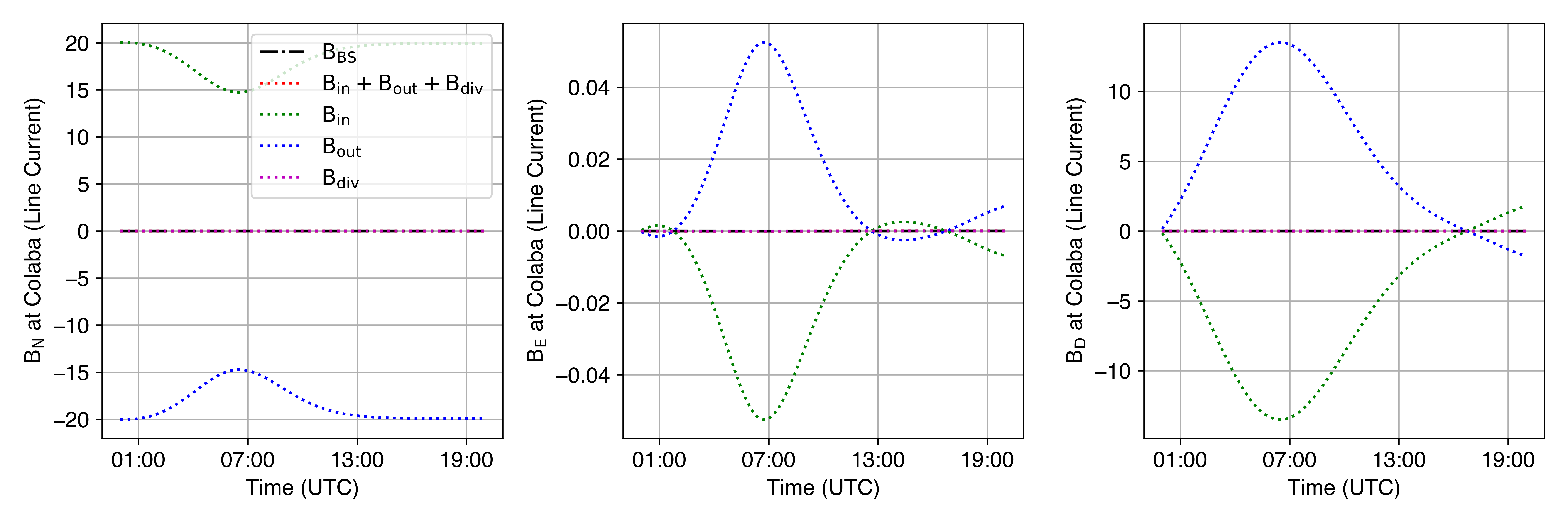}
\end{center}
\caption{$\bsint$, $\innerint$, $\diverr$, and $\outerr$ versus time derived from OpenGGCM simulation grid assuming infinite line current outside the simulation domain. Plots show $\mathbf{B}$ at Colaba, India. Same conclusions as in Figure~\ref{fig:BATSRUS line current}.}
\label{fig:OpenGGCM line current}
\end{figure}

\begin{figure}[!htb]
\begin{center}
    \includegraphics[width=\textwidth]{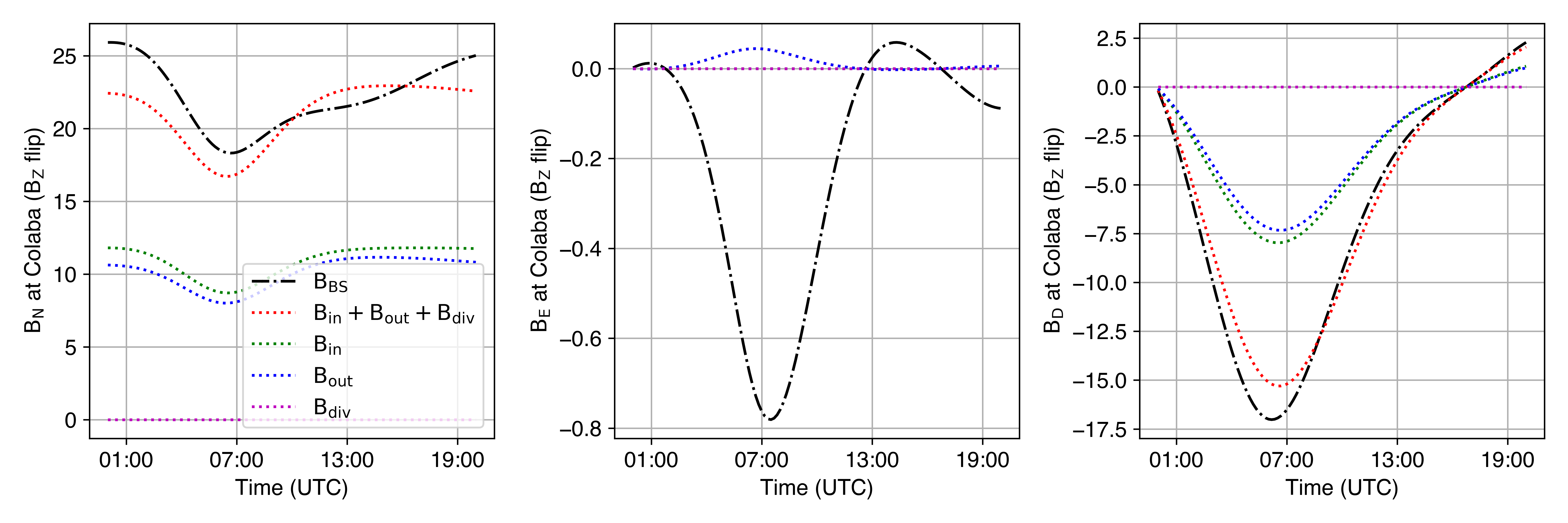}
\end{center}
\caption{$\bsint$, $\innerint$, $\diverr$, and $\outerr$ versus time derived from BATS--R--US simulation grid assuming a $B_Z^\text{\tiny{IMF}}$ flip within the simulation domain. Plots show $\mathbf{B}$ at Colaba, India. Since the magnetic field is discontinuous at $y=\pi$, we expect the Helmholtz Decomposition Theorem to be violated. Consistent with this, the red and black lines do not overlap.}
\label{fig:BATSRUS Bz flip}
\begin{center}
    \includegraphics[width=\textwidth]{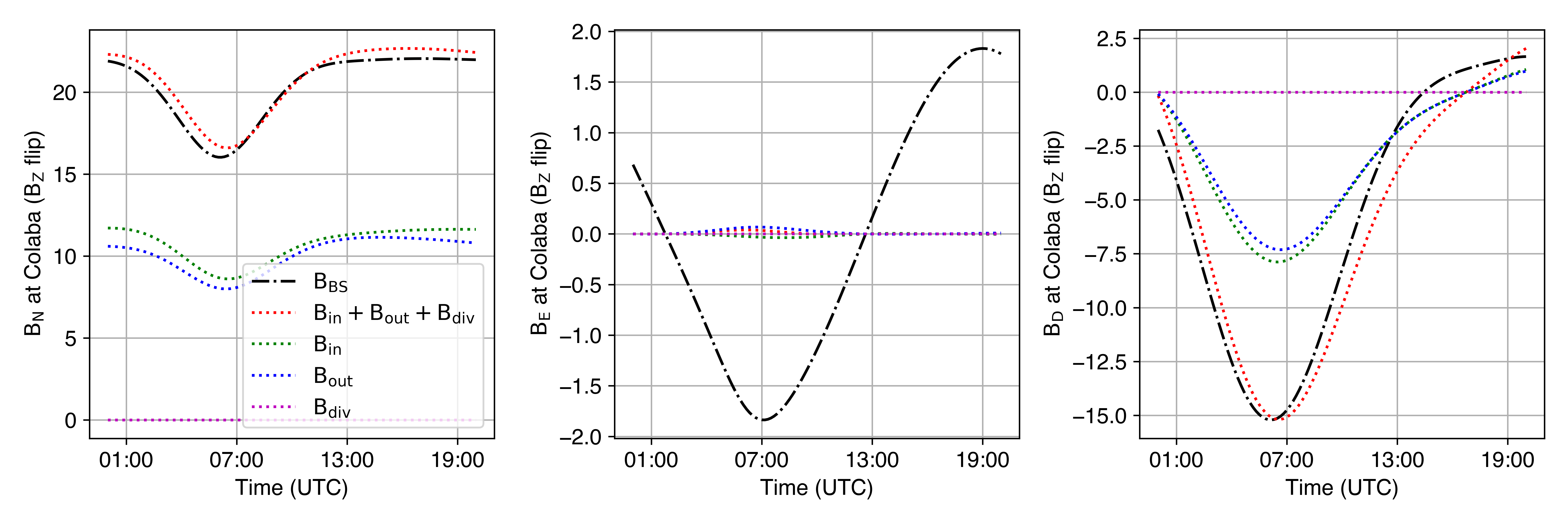}
\end{center}
\caption{$\bsint$, $\innerint$, $\diverr$, and $\outerr$ versus time derived from OpenGGCM simulation grid assuming a $B_Z^\text{\tiny{IMF}}$ flip within the simulation domain. Plots show $\mathbf{B}$ at Colaba, India. Same conclusions as in Figure~\ref{fig:BATSRUS Bz flip}.}
\label{fig:OpenGGCM Bz flip}
\end{figure}

\section{Additional Analysis}
\label{section:Additional }

We select five times for additional analysis across the surface of \remove[DT]{the }Earth. 01:00 and 04:00 at the beginning of the simulation, when the solar wind conditions are constant before the $B_Z^\text{\tiny{IMF}}$ flip. 07:00 shortly after the $B_Z^\text{\tiny{IMF}}$ flip, when we see significant shifts in $\mathbf{B}$. And 10:00 and 16:00, well after the $B_Z^\text{\tiny{IMF}}$ flip, when solar wind conditions are once again constant. We examine $\mathbf{B}$ across \remove[DT]{the }Earth at these times. In small increments in longitude and latitude, we step across \remove[DT]{the }Earth's surface, determining the values of the four integrals at each point. From these calculations, we generate heatmaps for the components of $\mathbf{B}$ across the surface of \remove[DT]{the }Earth. The results for $B_N$ are shown in Figures~\ref{fig:BATSRUS Bn heatmap} and \ref{fig:OpenGGCM Bn heatmap}. In the figures, each row is one of the four integrals, and each column is one of the selected times. Results for $B_N$ and $B_D$ are shown in the appendix.

The previously observed trends continue to be valid. In Figures~\ref{fig:BATSRUS Bn heatmap} and \ref{fig:OpenGGCM Bn heatmap}, before the $B_Z^\text{\tiny{IMF}}$ flip at 06:00, we see small contributions from all four integrals as demonstrated by the similar color intensities in all four rows in the leftmost two columns. After the $B_Z^\text{\tiny{IMF}}$ flip, $\innerint$ is the dominant contributor to $\bsint$, as seen in the similar color shading in the top two rows in the rightmost two columns. $\diverr$ makes larger contributions for BATS--R--US than for OpenGGCM. These conclusions are also the same for $B_E$ and $B_D$ heatmaps, which are in the appendix. 

These trends are highlighted in Figures~\ref{fig:BATSRUS average term} and \ref{fig:OpenGGCM average term}.  They show the surface average of $\diverr$ and $\outerr$ as a fraction of the surface average of $\bsint$ for North, East, and Down components. The surface averages are calculated over \remove[DT]{the }Earth's surface using the heatmap data at the times selected above.  The fractions are smaller at large $|\bsint|$.  BATS--R--US fractions are typically larger than OpenGGCM fractions.

\remove[DT]{The fact that the BATS-R-US and OpenGGCM results are similar, but not identical is illustrated in Figure~\ref{fig:BATSRUS OpenGGCM comparison}.  The figure compares OpenGGCM to BATS--R--US surface-average $\bsint$ estimates.  If the simulations provided identical results, the dots would form a 45-degree diagonal line.  While the results are not identical, the two simulations give similar results with better agreement at large $|\bsint|$.}

To understand correlations between $\bsint$, $\innerint$, $\diverr$, and $\outerr$, we plot the heatmap data in one more way. In Figures~\ref{fig:BATS--R--US Bn integrals vs Biot} and \ref{fig:OpenGGCM Bn integrals vs Biot}, we plot the $\innerint$, $\diverr$, and $\outerr$ versus $\bsint$. Like the earlier figures, the rows are one of the three integrals versus $\bsint$, and the columns are the times selected above. The horizontal and vertical scales in each column are identical. An upward diagonal 45--degree line indicates that the applicable integral and $\bsint$ are correlated. A downward 45--degree diagonal line indicates that they are anti--correlated. A horizontal line indicates that the integral makes a small contribution to $\bsint$. 

\begin{figure}[!htb]
\includegraphics[width=\textwidth]{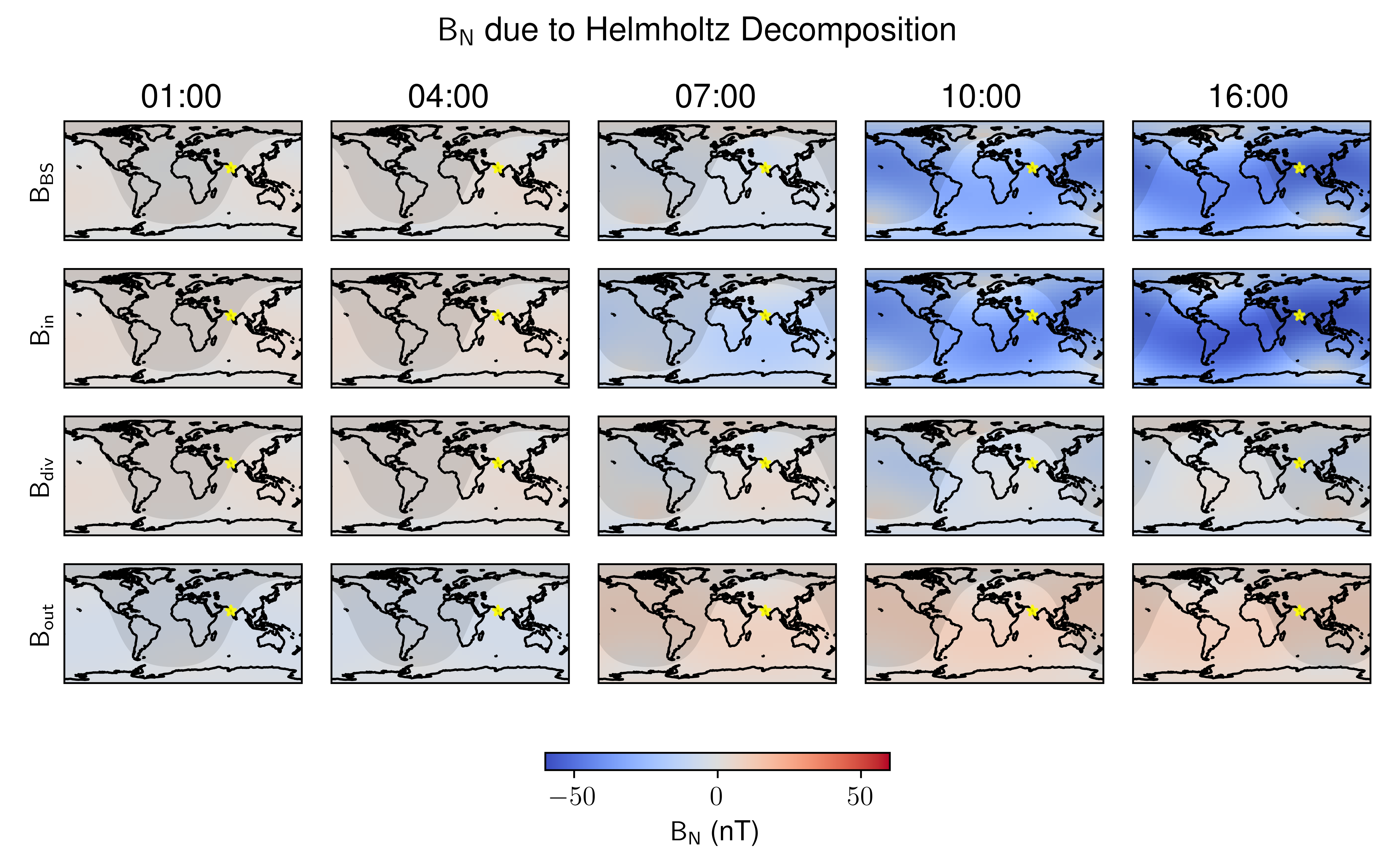}
\caption{$B_N$ contributions from $\bsint$, $\innerint$, $\diverr$, and $\outerr$ integrals derived from BATS--R--US simulation. Each row is one of the four integrals. Each column is one of the five times. Color designates $B_N$ at each point.  Before 06:00 $B_Z^\text{\tiny{IMF}}$ flip, all contributions are small.  As the simulation progresses after $B_Z^\text{\tiny{IMF}}$ flip, $\innerint$ becomes the dominant contributor to $\bsint$.}
\label{fig:BATSRUS Bn heatmap}
\vspace{5mm}
\includegraphics[width=\textwidth]{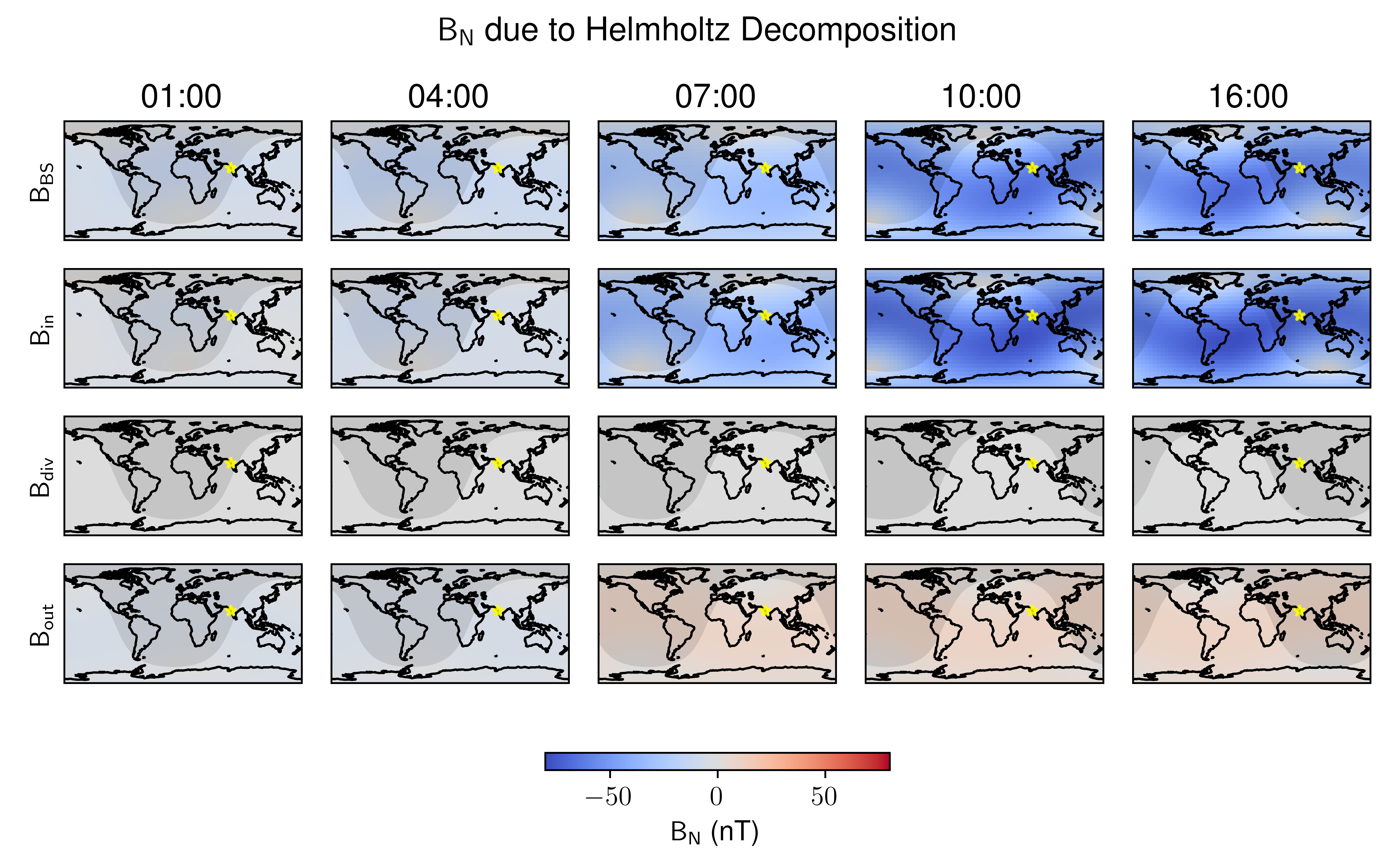}
\caption{$B_N$ from $\bsint$, $\innerint$, $\diverr$, and $\outerr$ integrals derived from the OpenGGCM simulation across \remove[DT]{the }Earth's surface. Same format and conclusions as Figure~\ref{fig:BATSRUS Bn heatmap}.}
\label{fig:OpenGGCM Bn heatmap}
\end{figure}

\begin{figure}[!htb]
\includegraphics[width=0.9\textwidth]{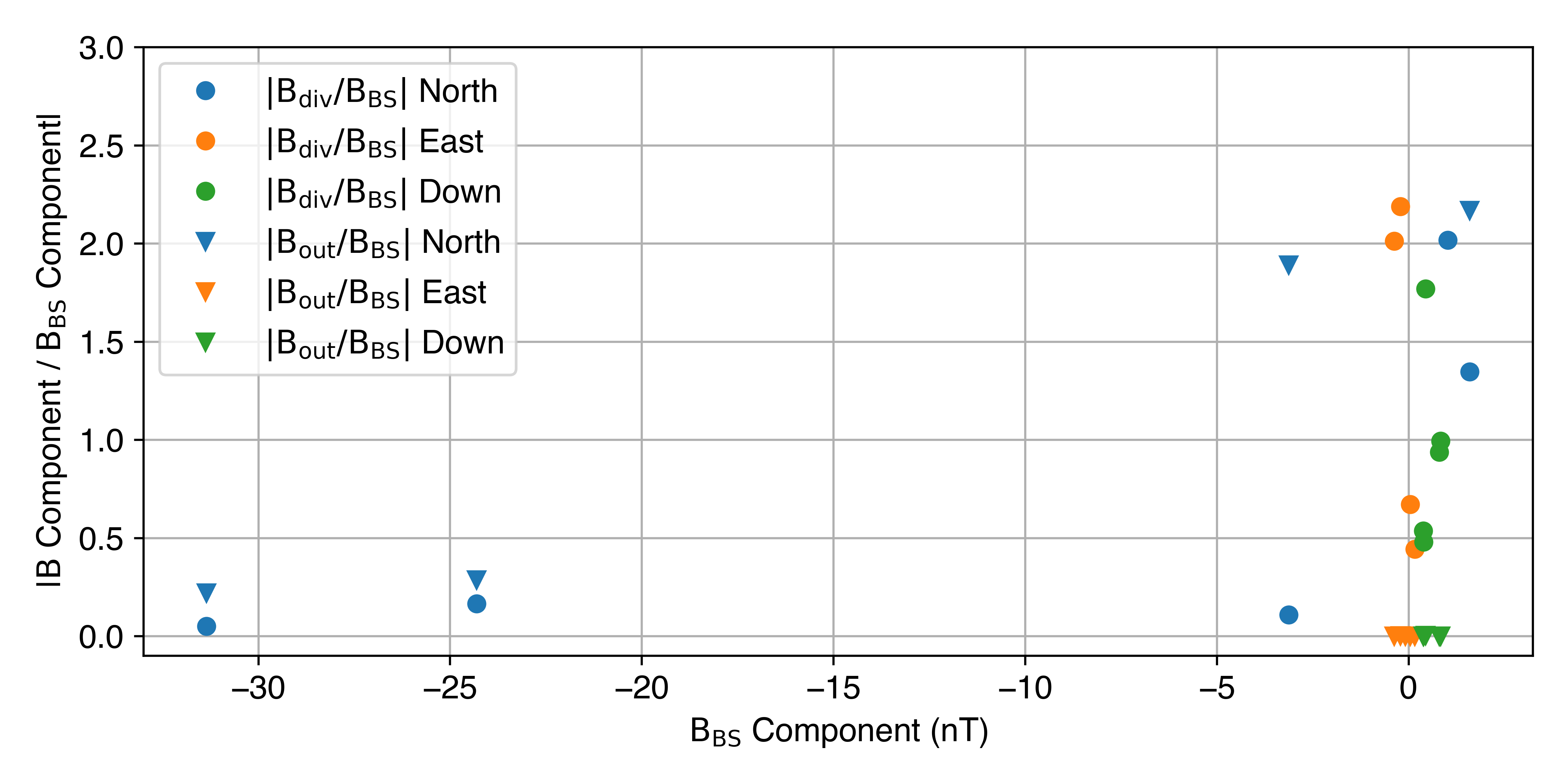}
\caption{BATS-R-US $\diverr$ and $\outerr$ as fraction of $\bsint$ for North, East, and Down (e.g., $|\text{B}_{\text{div}}(\text{North})/{\text{B}_\text{BS}}(\text{North})|$). The surface averages are calculated using the BATS-R-US heatmap (Figure~\ref{fig:BATSRUS Bn heatmap}) and the fractions are smaller at large $|\bsint|$.  BATS--R--US fractions are generally larger than OpenGGCM fractions in Figure~\ref{fig:OpenGGCM average term}}.
\label{fig:BATSRUS average term}
\end{figure}

\begin{figure}[!htb]
\includegraphics[width=0.9\textwidth]{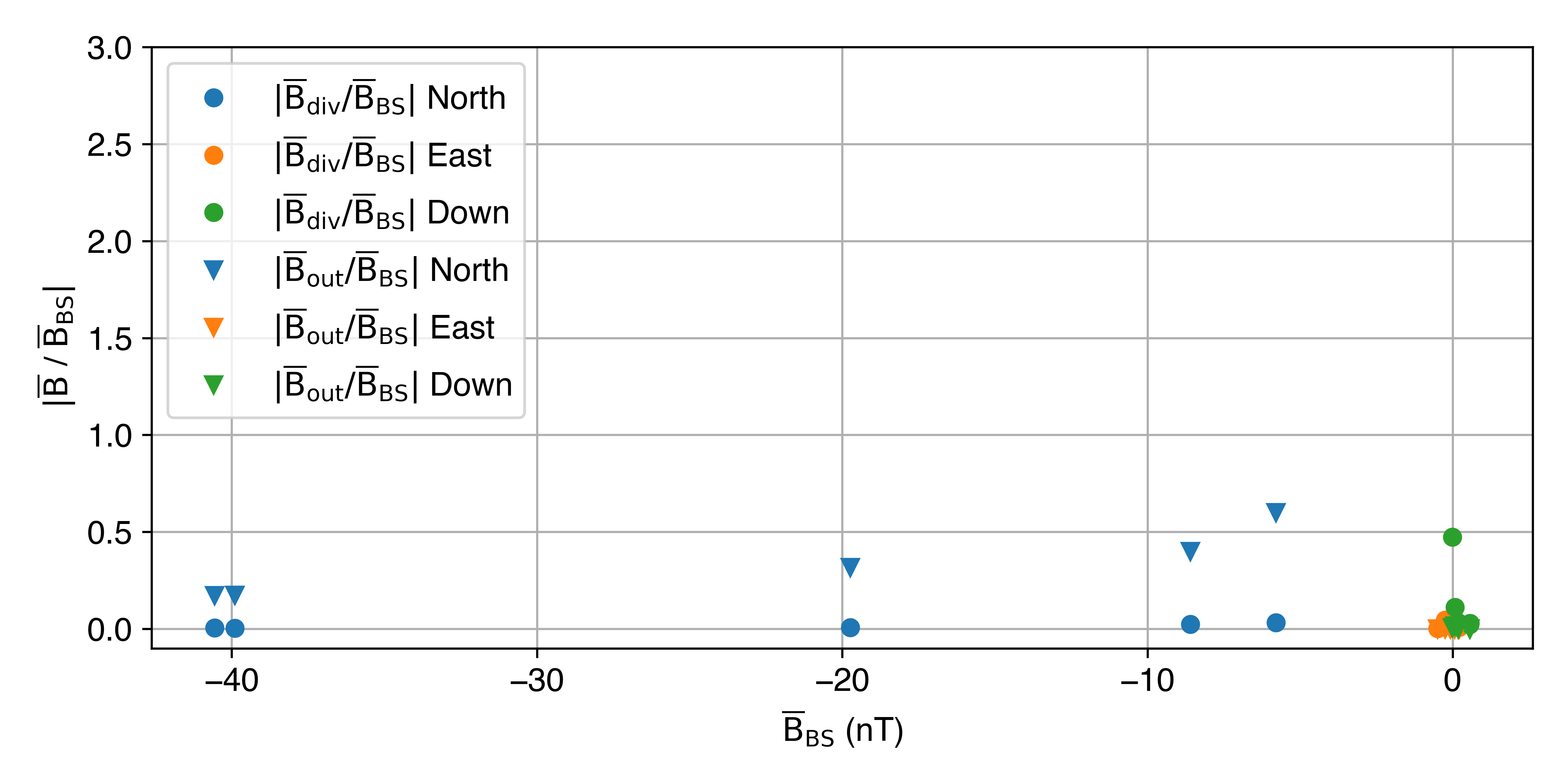}
\caption{OpenGGCM $\diverr$ and $\outerr$ as fraction of $\bsint$ for North, East, and Down. The surface averages are calculated using the OpenGGCM heatmap (Figure~\ref{fig:OpenGGCM Bn heatmap}) and the fractions are smaller at large $|\bsint|$.}
\label{fig:OpenGGCM average term}
\end{figure}
 

\change[DT]{Ideally, to}{To} justify the use of the Biot--Savart Law in computing the magnetic field due to magnetospheric currents, $\innerint$ versus $\bsint$ plots would be a 45-degree diagonal line with the $\innerint$ approximately equal to $\bsint$ at all points. $\diverr$ versus $\bsint$ would be a horizontal line with the $\diverr$ being close to 0. And $\outerr$ versus $\bsint$ would show $\outerr$ being much smaller than $\bsint$, and close to a horizontal line. 

The OpenGGCM plots most closely match these \remove[DT]{desired} characteristics. We also observe that the results show some anti-correlated behavior. $\outerr$ is anti-correlated for BATS--R--US throughout the simulation, and the $\outerr$ is anti--correlated for OpenGGCM after the $B_Z^\text{\tiny{IMF}}$ flip.

We note that the $\diverr$ and $\outerr$ are spatially correlated. In Figures~\ref{fig:BATS--R--US Bn integrals vs Biot} and \ref{fig:OpenGGCM Bn integrals vs Biot}, the results are color--coded according to the geolatitude on \remove[DT]{the }Earth's surface. While not shown, we see similar results when geolongitude is used to color code the data. Since $\mathbf{B}$ at two nearby points on Earth are correlated, it is expected that the $\diverr$ and $\outerr$ also are correlated. We see similar correlations for $B_E$ and $B_D$ in the appendix.

\begin{figure}[!htb]
\includegraphics[width=\textwidth]{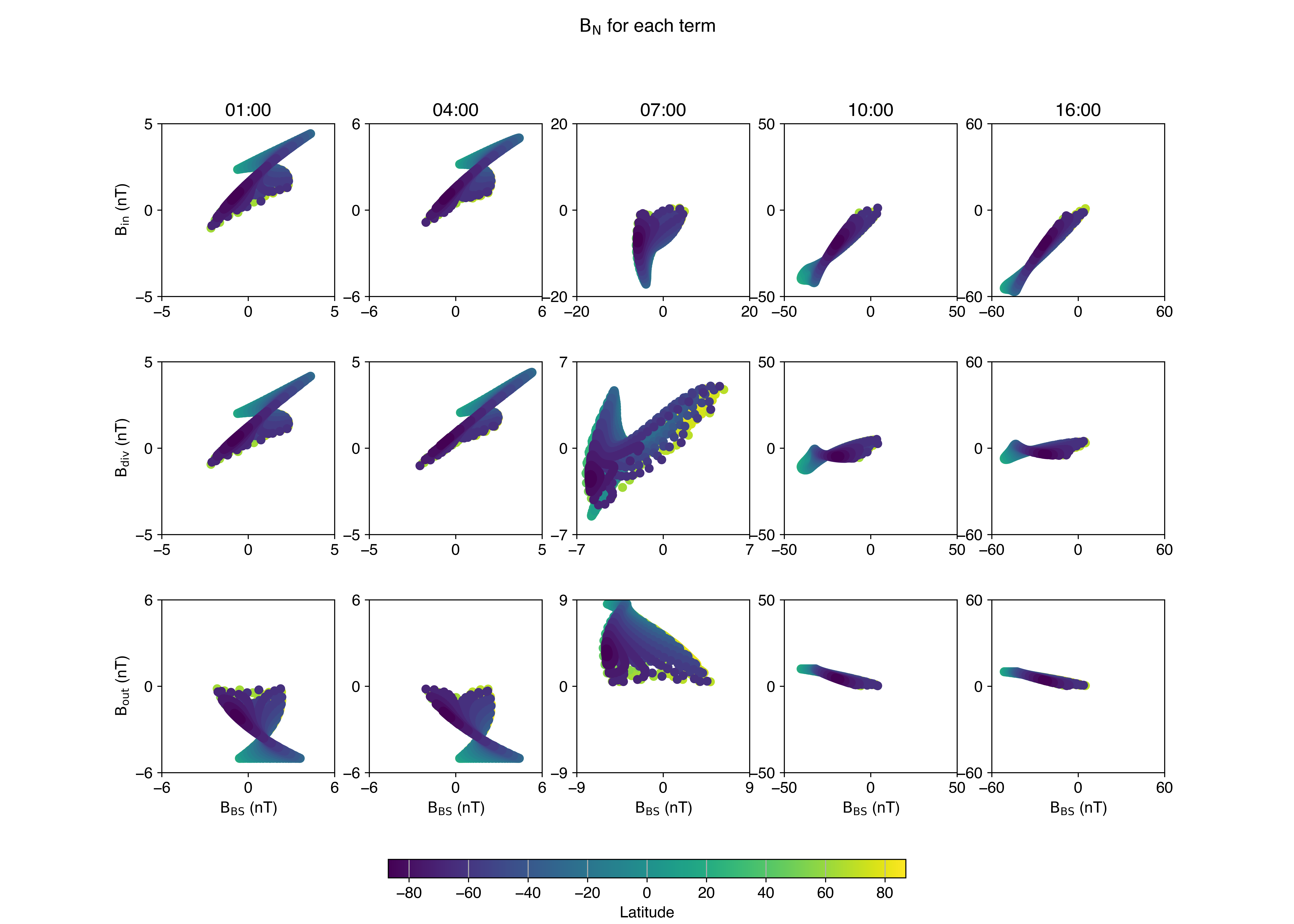}
\caption{$B_N$ from $\innerint$, $\diverr$, and $\outerr$ versus $\bsint$ using BATS--R--US results in Figure~\ref{fig:BATSRUS Bn heatmap}. Each row is one of the three integrals versus Biot--Savart. Each column is one of the selected times.  Plots illustrate integrals are spatially correlated.}
\label{fig:BATS--R--US Bn integrals vs Biot}
\includegraphics[width=\textwidth]{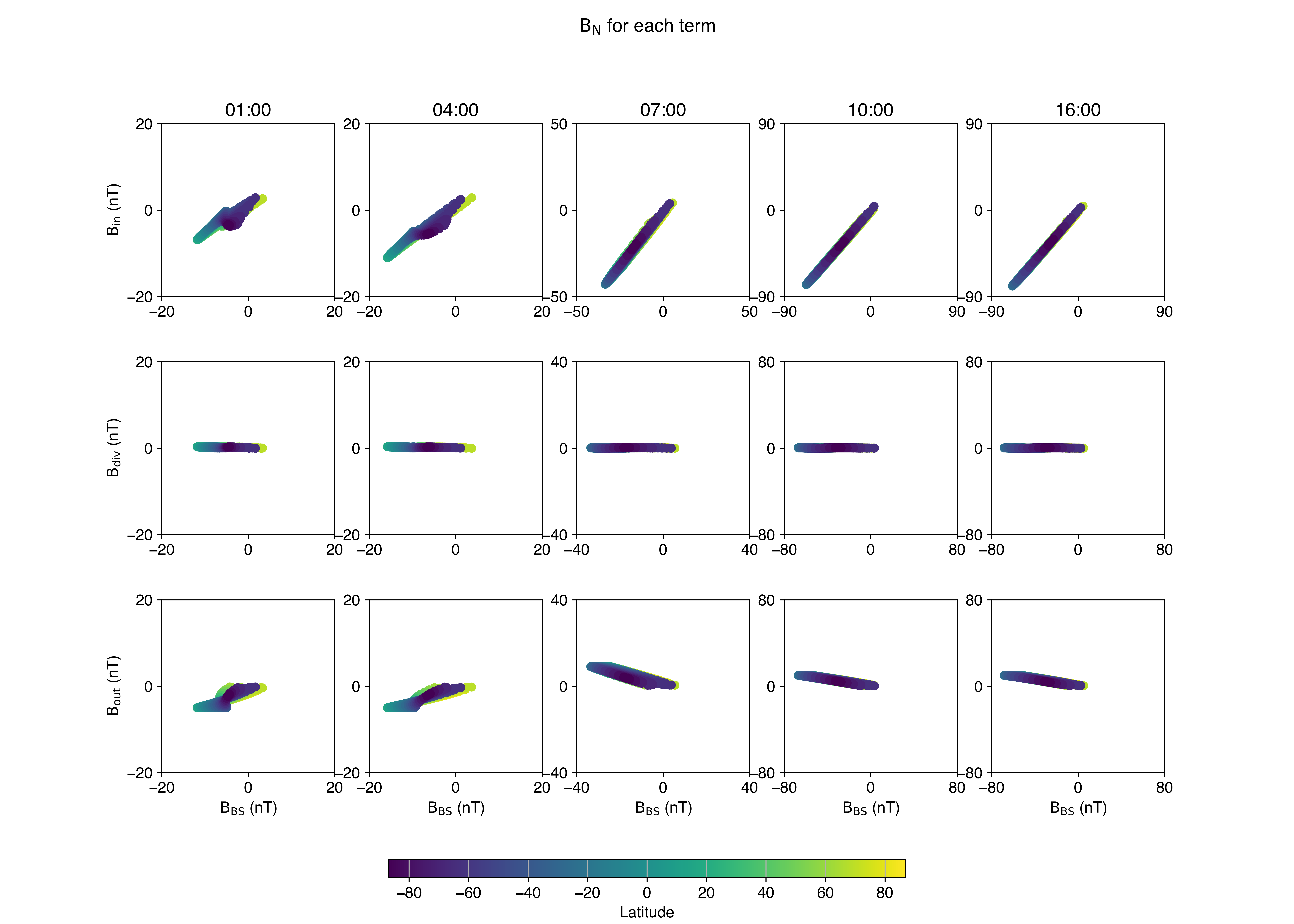}
\caption{$B_N$ from $\innerint$, $\diverr$, and $\outerr$ versus $\bsint$ using OpenGGCM results in Figure~\ref{fig:OpenGGCM Bn heatmap}. Same format and similar conclusions as Figure~\ref{fig:BATS--R--US Bn integrals vs Biot}.}
\label{fig:OpenGGCM Bn integrals vs Biot}
\end{figure}

\clearpage
\section{Comparison to Superstorm}

We compare the results above to those we have for a superstorm, and the results are similar. \citeA{Thomas2024} examined the Carrington event using an SWMF simulation \cite{Blake_2021, Ngwira_2014}. Figure~\ref{fig:Carrington average term} has the same format as Figures~\ref{fig:BATSRUS average term} and \ref{fig:OpenGGCM average term}.  Following \citeA{Thomas2024}, the quantities in the figure are calculated at 05:00 and 06:00 in the early stages of the storm, 06:30 at storm peak, and 07:00 and 08:00 during storm recovery. The Carrington simulation results are similar to those discussed in section~\ref{section:Magnitude term}. The fractions are smaller at large $|\bsint|$.  

\begin{figure}[!htb]
\includegraphics[width=\textwidth]{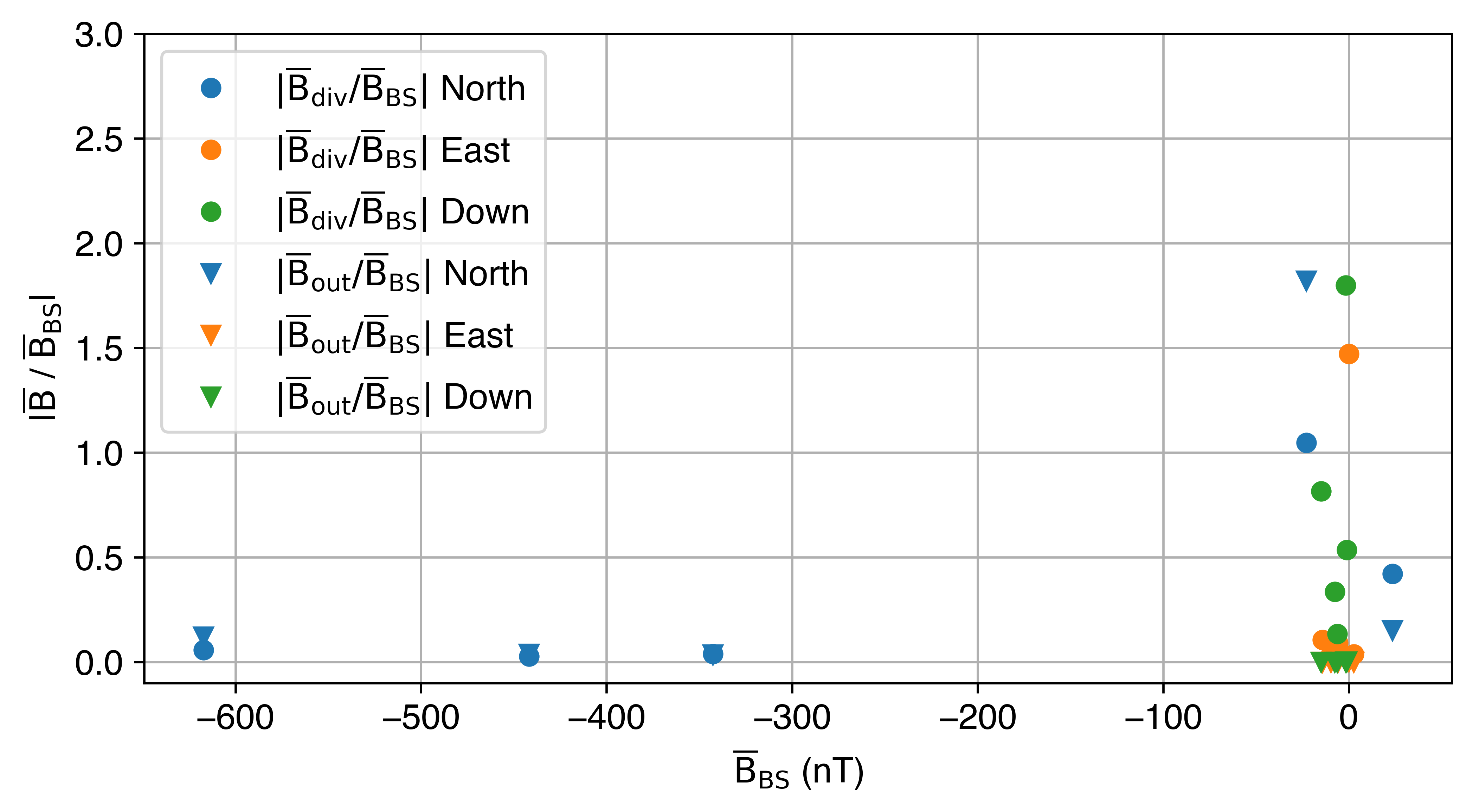}
\caption{For the Carrington simulation, BATS--R--US $\diverr$ and $\outerr$ as fraction of $\bsint$ for North, East, and Down.  Same format and conclusions as Figures~\ref{fig:BATSRUS average term} and \ref{fig:OpenGGCM average term}}
\label{fig:Carrington average term}
\end{figure}

\section{Conclusions}

\add[DT]{Our analysis shows that using the traditional Biot--Savart approach to determine magnetospheric contributions to the magnetic field at a point on Earth has issues that can be avoided.  Numerically, it is faster to calculate $\innerint$, the surface integral over the inner boundary, rather than calculating $\bsint$, the integral across the magnetospheric volume.  Furthermore, $\innerint$ is independent of $\divB$ and the outer boundary of the magnetosphere. Using $\innerint$ is an approach implemented, for example, in SWMF as described in \citeA{Gombosi2021}.}

Our analysis examined MHD simulations, where we observed two different results for $\diverr$ and $\outerr$. For large $|\bsint|$ computed using equation~\ref{eqn:biotsimple}, the surface-averaged:
    \begin{itemize}
    \item $\outerr$ is $20-30$\% of $\bsint$ for BATS--R--US and ${\sim20}$\% for OpenGGCM.
    \item $\diverr$ is $5-20$\% of $\bsint$ for BATS--R--US and $<1$\% for OpenGGCM.
    \end{itemize}
These values are derived from the heatmap surface averages at 10:00 and 16:00 in Figures~\ref{fig:BATSRUS Bn heatmap} and \ref{fig:OpenGGCM Bn heatmap}. For small $|\bsint|$, the percentages are larger, with BATS--R--US having the largest percentages. This conclusion is from the heatmap surface averages before 07:00.  These conclusions are consistent with those seen in a simulation of a superstorm \cite{Thomas2024}.  

This result implies that using only the Biot-Savart volume integral to compute the magnetic field on Earth's surface due to magnetospheric currents has two uncertainties that should be acknowledged (1) An approximation uncertainty due to the omission of the contribution of currents outside of the magnetospheric volume, which are captured by $\outerr$ and (2) A self--consistency uncertainty associated with $\diverr$. Even if $\outerr$ is accounted for, to be consistent with the Helmholtz decomposition theorem, the contribution from $\diverr$ must be included.

There is a third ``model uncertainty" that should be acknowledged when interpreting the magnetic field on Earth's surface estimated using MHD simulation results.  We note the substantial differences in the results from different MHD models. Different simulations provide similar $\mathbf{B}$ trends (i.e., similarly shaped curves), but the specific values of $\mathbf{B}$ substantially differ. In the conditions we examined, differences of up to a factor of \mbox{$\approx2$} occur. These differences are expected because different MHD models use different methods for \remove[DT]{handling $\divB$ numerical errors,} solving the MHD equations and modeling the ionosphere.

These conclusions indicate that caution should be used in interpreting MHD results based on Biot--Savart analysis.  An analysis that uses the Biot--Savart Law to compute $\mathbf{B}$ at a point using magnetospheric current density estimates may have substantial uncertainty.  We should not assert that the actual values based on an MHD simulation are accurate, but we can trust similar trends across models.  Concern is warranted when $|\bsint|$ is small and the contributions from {\divB} and outer surface boundary integrals are large compared to the magnetospheric contributions.  

If $\outerr$ is a concern, the analysis can include the outer surface integral in the Biot--Savart analysis.  Or the analysis can use the inner surface integral through equation~\ref{eqn:biot G}.  \add[DT]{We also note that $\outerr$ is over the outer boundary of the simulation domain. Thus, it is affected by the dimensions of the domain, and this can account for differences between simulations.}

Regarding $\diverr$, we see differences between the simulations examined.  It is already minimized in OpenGGCM.  For BATS--R--US, the eight-wave method \change[DT]{has a known limitation: $\diverr$ may grow when the {\divB} source is large}{calculates $\divB$ at the truncation level, but $\divB$ may be significant near discontinuities, such as shocks} \cite{Gombosi2003adaptive}. As with $\outerr$, analyses can employ the inner surface integral through equation~\ref{eqn:biot G} to account for $\diverr$. 

We note that the $\diverr$ and $\outerr$ are spatially correlated.  This observation is consistent with the known behavior of $\mathbf{B}$ on \remove[DT]{the }Earth's surface.  

Overall, we believe the uncertainties described above are reasonable bounds for contributions from the {\divB} and outer boundary integrals. As space weather researchers, we are interested in what happens in an evolving scenario when $|\bsint|$ is large, not a stagnant one when it is small. And as \citeA{Ngwira_2014} observed, solar wind conditions employed in MHD simulations require realistic variability at the temporal scales observed in real events. Solar wind variability is needed to balance fluctuations in the magnetospheric structures under extreme driving conditions. Uncertainties in the range $5-30$\% of $\bsint$ are expected. 

\add[DT]{Using $\innerint$ has fewer limitations and is computationally more efficient. Therefore, we recommend it over a Biot--Savart approach.}

\remove[DT]{Comparing these to the inherent variability between MHD simulations, we conclude that using the Biot--Savart Law to estimate the magnetic field at magnetometer sites is reasonable.   However, we also acknowledge the known limitations of the Biot--Savart approach.  It is ineffective when treating strong, non-symmetric, or time-dependent situations .  And MHD results demonstrate substantial signal propagation times.}

\section*{Open Research Section}
The software used in this analysis can be found at \citeA{Thomas_Github}.  The SWMF dataset is available at \citeA{SWMF_Bz}, and the OpenGGCM dataset is available at \citeA{OpenGGCM_Bz}.  The Carrington superstorm data set is available at \citeA{Blake_Carrington}.

\section*{Conflict of Interest Declaration}
\add[DT]{
The authors declare there are no conflicts of interest for this manuscript.
}

\acknowledgments

This work was partially supported by NASA Grant 80NSSC20K0589.

This work was carried out using the SWMF and BATS--R--US tools developed at the University of Michigan’s Center for Space Environment Modeling (CSEM). The modeling tools are available through the University of Michigan for download under a user license; an open--source version is available at \cite{swmfsoftware}.

This work also used OpenGGCM, which is housed at the Space Science Center of the University of New Hampshire. More detailed information about the OpenGGCM can be found on the OpenGGCM wiki page \cite{openggcmsoftware} and on the NASA/CCMC page \cite{ccmcsoftware} where the OpenGGCM is available as a community model for runs on demand. We also thank Jimmy Raeder for feedback on the manuscript.

\add[DT]{Both SWMF and OpenGGCM are available at} \change[DT]{The}{the} Community Coordinated Modeling Center (CCMC)\add[DT]{. CCMC} executed the simulations at Goddard Space Flight Center (Run IDs Bob\_Weigel\_070323\_3 and Dean\_Thomas\_052924\_1).

\appendix
\section{Additional Figures}
As discussed in section~\ref{section:Magnitude term} of the report, the conclusions drawn from the North, East, and Down components of $\mathbf{B}$ are consistent. Previously, we only discussed $B_N$. For completeness, we present the figures for $B_E$ and $B_D$.

\begin{figure}[!htb]
\includegraphics[width=0.9\textwidth]{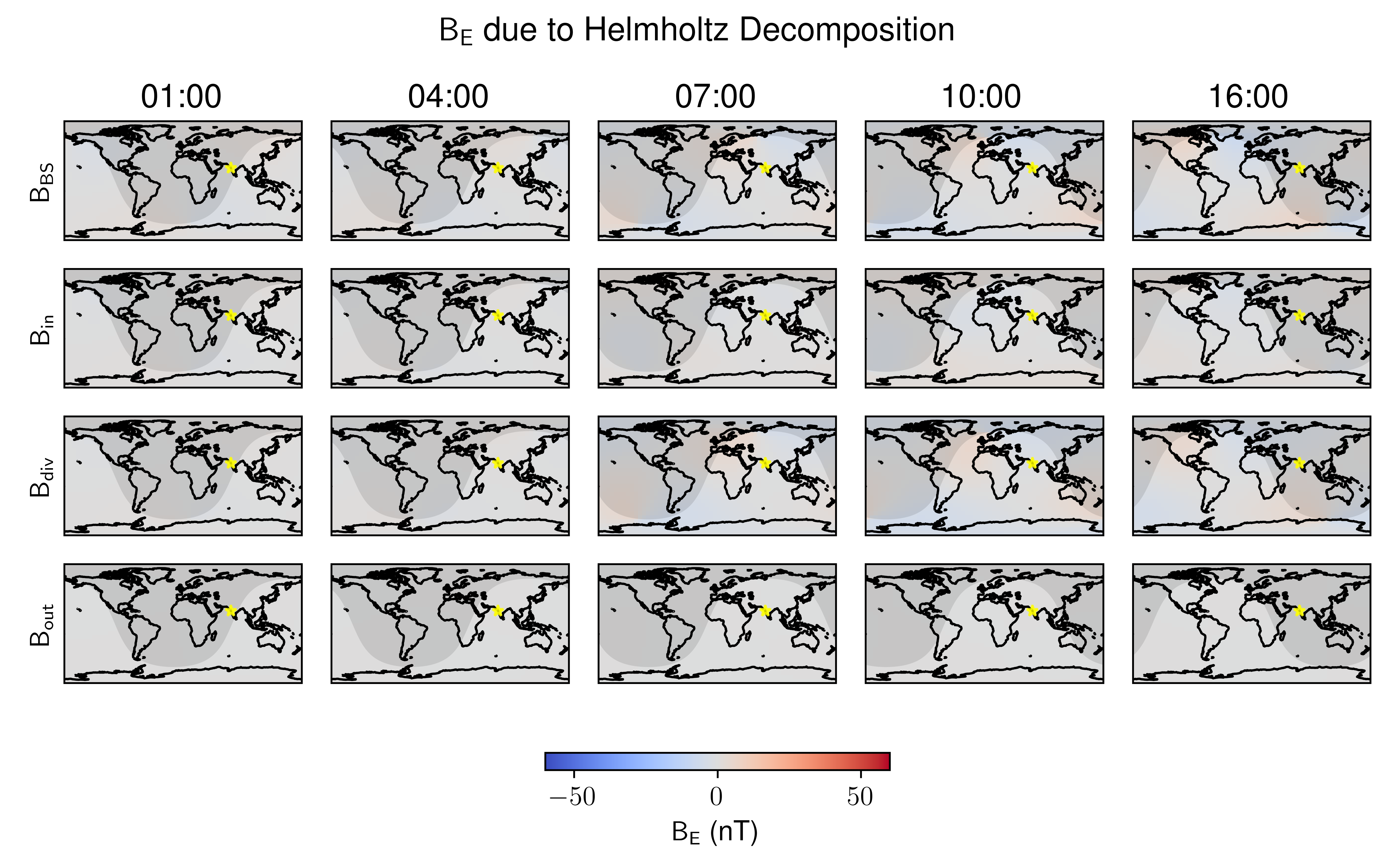}
\caption{$B_E$ from $\bsint$, $\diverr$, $\innerint$ and $\outerr$ derived from BATS--R--US simulation. Same format as Figure~\ref{fig:BATSRUS Bn heatmap}}
\label{fig:BATSRUS Be heatmap}.
\vspace{5mm}
\includegraphics[width=0.9\textwidth]{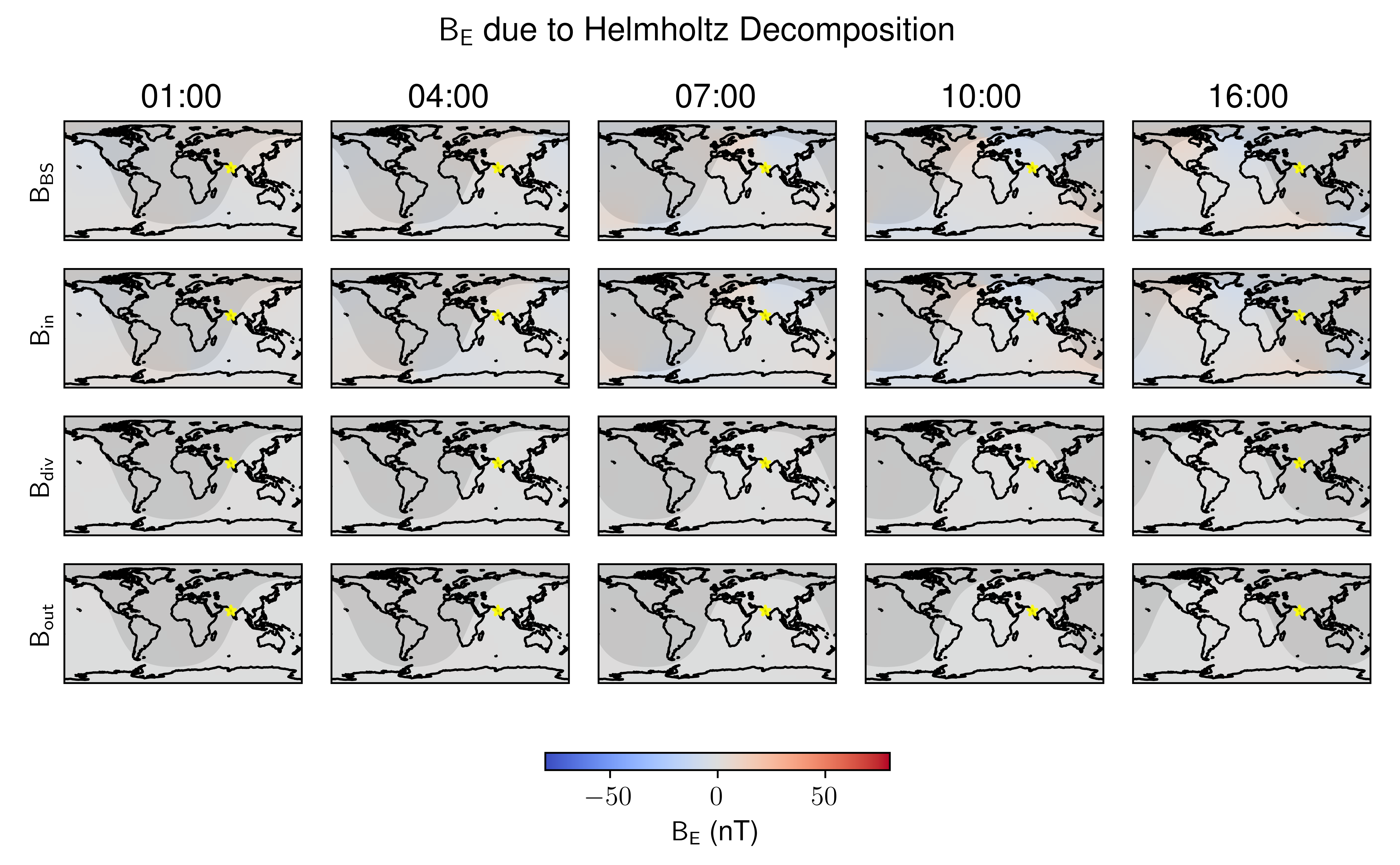}
\caption{$B_E$ from $\bsint$, $\diverr$, $\innerint$ and $\outerr$ derived from OpenGGCM simulation. Same format as Figure~\ref{fig:OpenGGCM Bn heatmap}}
\label{fig:OpenGGCM Be heatmap}.
\end{figure}

\begin{figure}[!htb]
\includegraphics[width=0.9\textwidth]{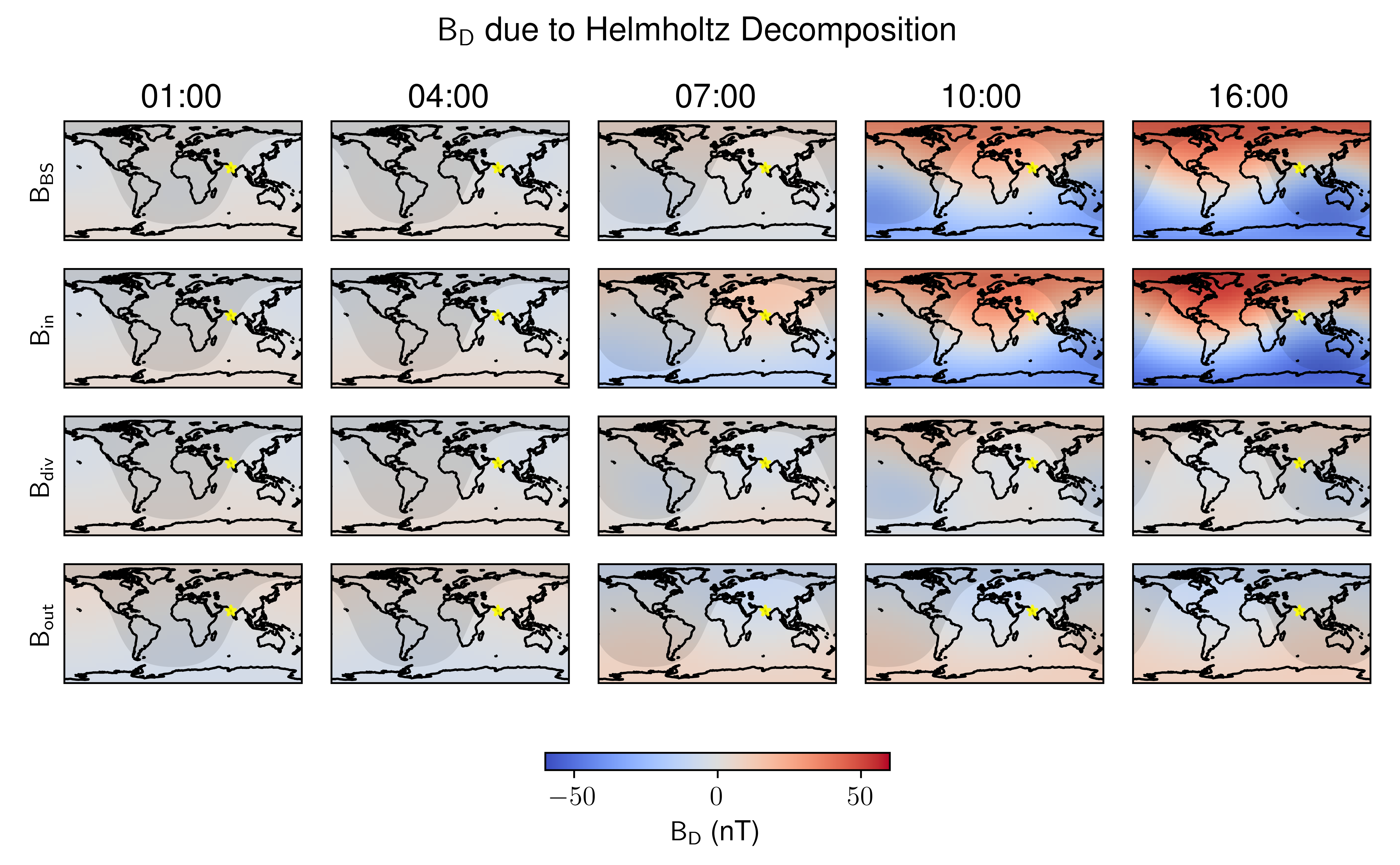}
\caption{$B_D$ from $\bsint$, $\diverr$, $\innerint$ and $\outerr$ derived from BATS--R--US simulation. Same format as Figure~\ref{fig:BATSRUS Bn heatmap}}
\label{fig:BATSRUS Bd heatmap}.
\vspace{5mm}
\includegraphics[width=0.9\textwidth]{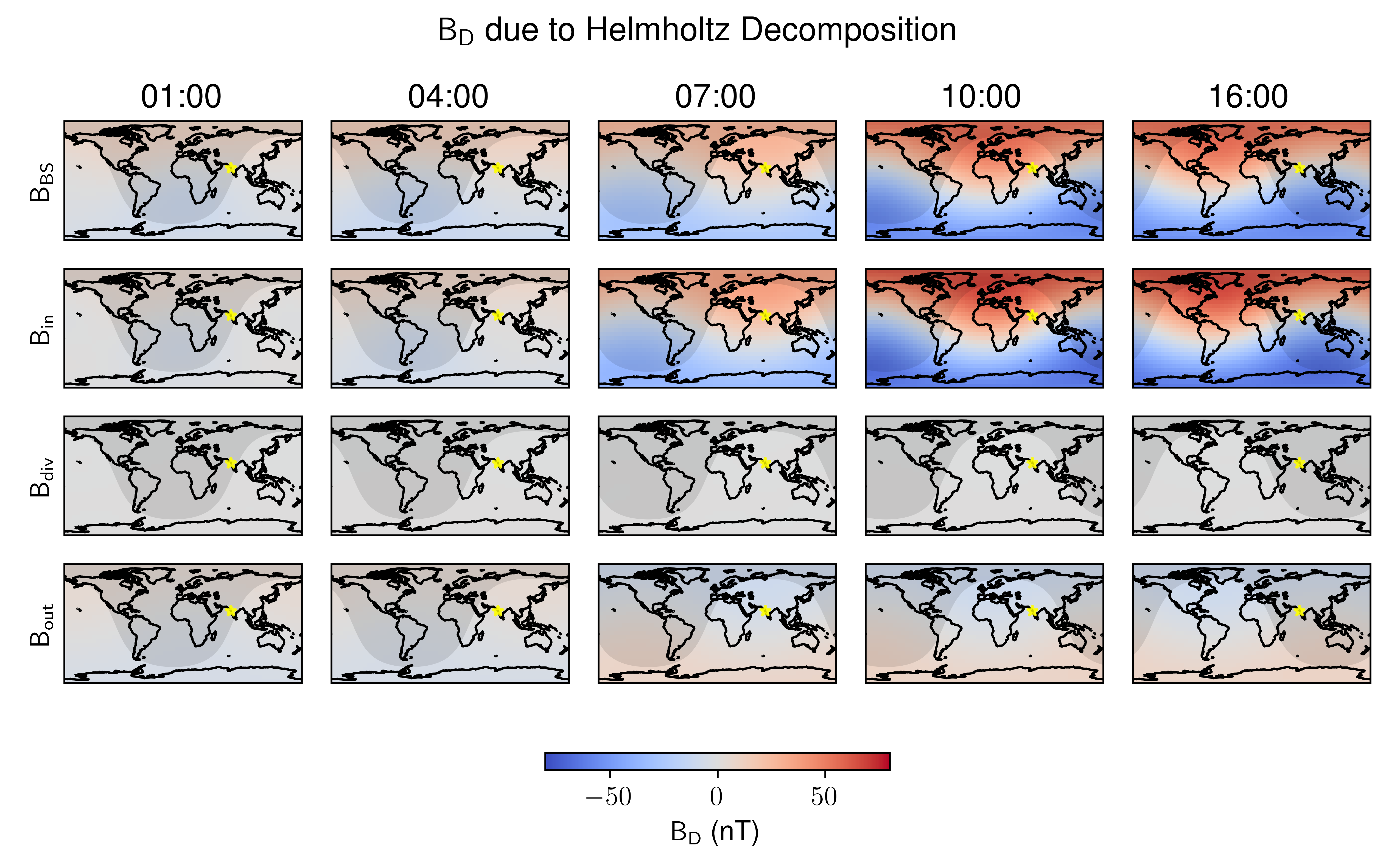}
\caption{$B_D$ from $\bsint$, $\diverr$, $\innerint$ and $\outerr$ derived from OpenGGCM simulation. Same format as Figure~\ref{fig:OpenGGCM Bn heatmap}}.
\label{fig:OpenGGCM Bd heatmap}
\end{figure}

\begin{figure}[!htb]
\includegraphics[width=0.9\textwidth]{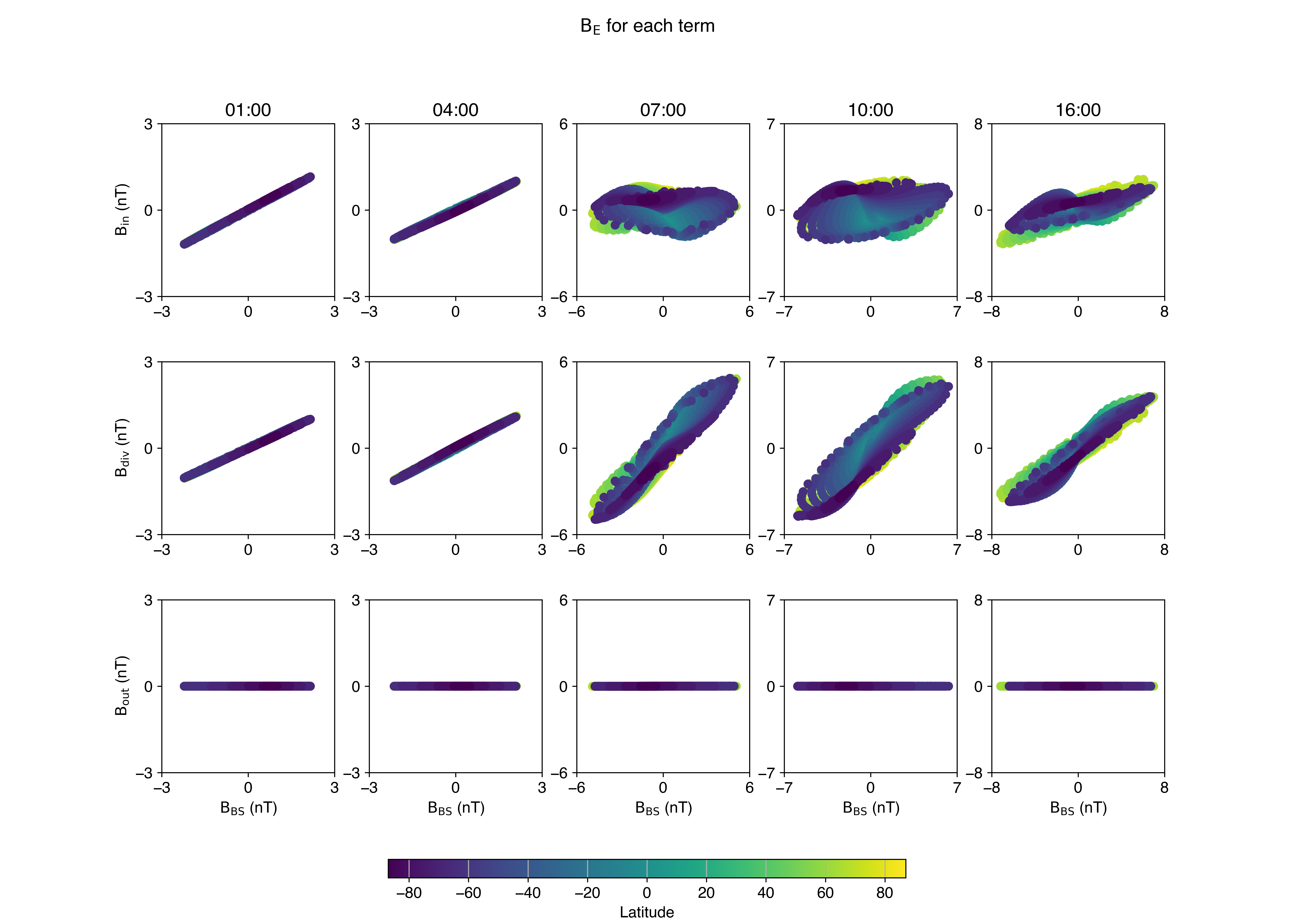}
\caption{$B_E$ from $\innerint$, $\diverr$, and $\outerr$ versus $\bsint$ using BATS--R--US data. Same format as Figure~\ref{fig:BATS--R--US Bn integrals vs Biot}}.
\label{fig:BATS--R--US Be integrals vs Biot}
\vspace{5mm}
\includegraphics[width=0.9\textwidth]{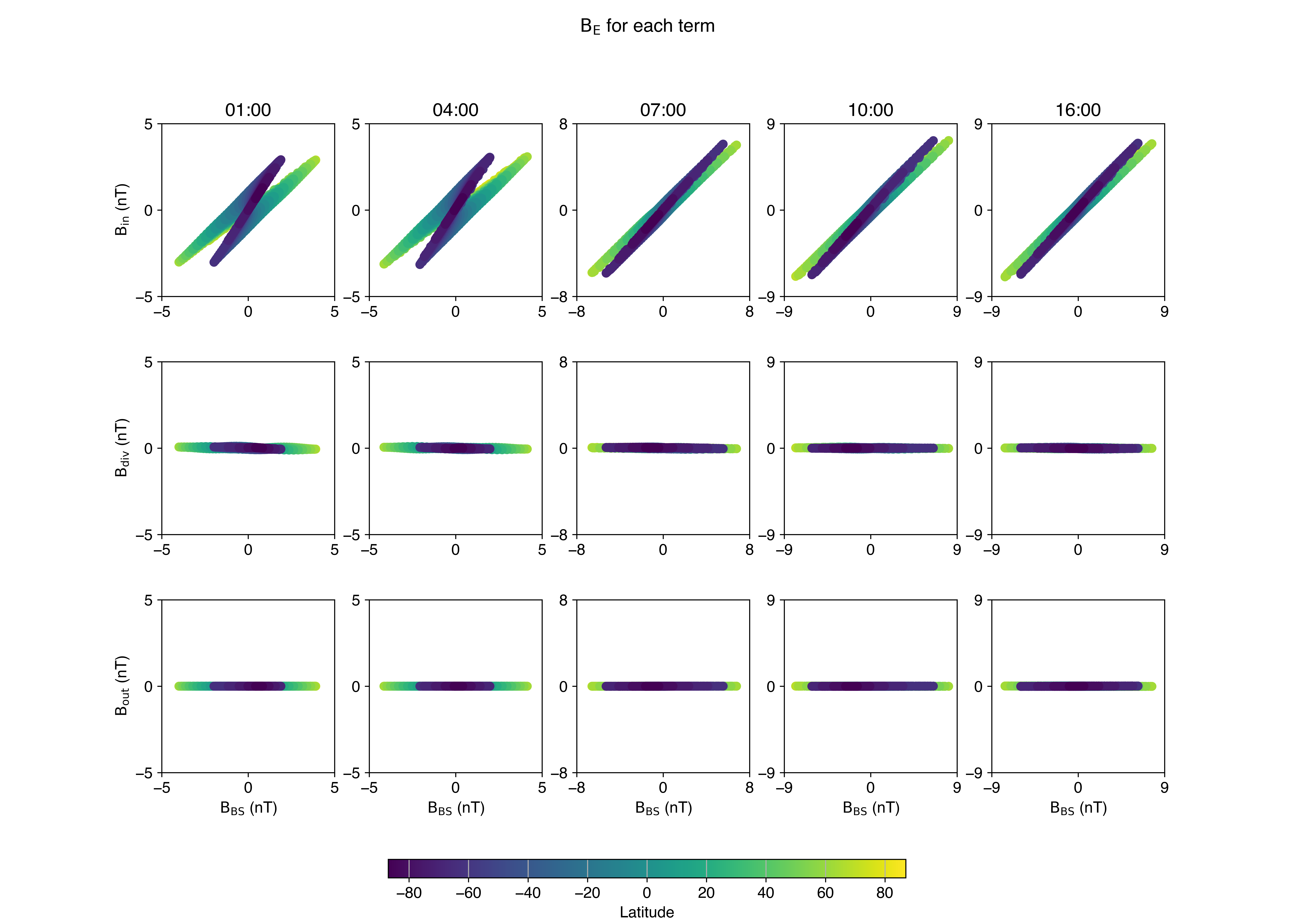}
\caption{$B_E$ from $\innerint$, $\diverr$, and $\outerr$ versus $\bsint$ using OpenGGCM data. Same format as Figure~\ref{fig:OpenGGCM Bn integrals vs Biot}}.
\label{fig:OpenGGCM Be integrals vs Biot}
\end{figure}

\begin{figure}[!htb]
\includegraphics[width=0.9\textwidth]{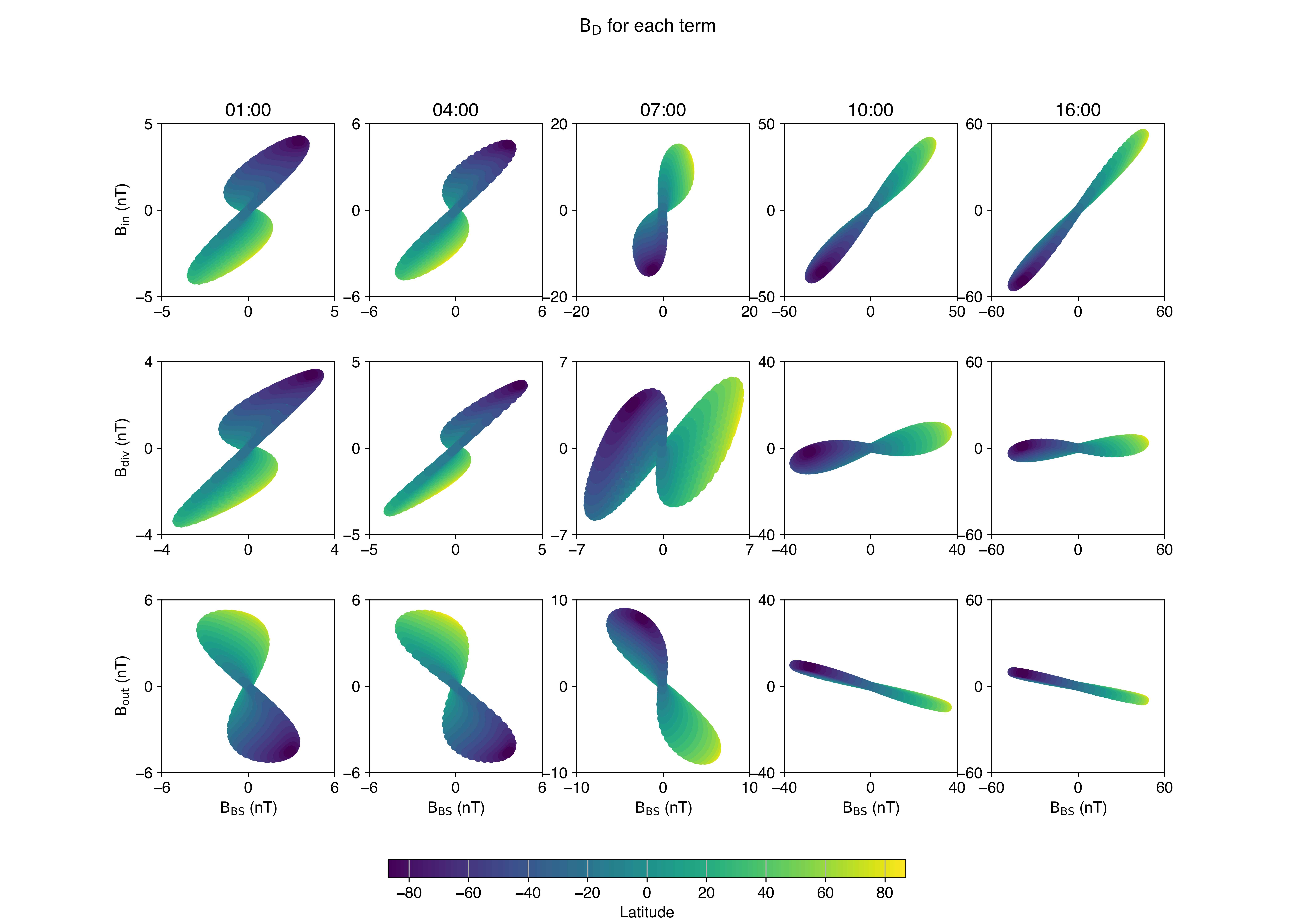}
\caption{$B_D$ from $\innerint$, $\diverr$, and $\outerr$ versus $\bsint$ using BATS--R--US data. Same format as Figure~\ref{fig:BATS--R--US Bn integrals vs Biot}}.
\label{fig:BATS--R--US Bd integrals vs Biot}
\vspace{5mm}
\includegraphics[width=0.9\textwidth]{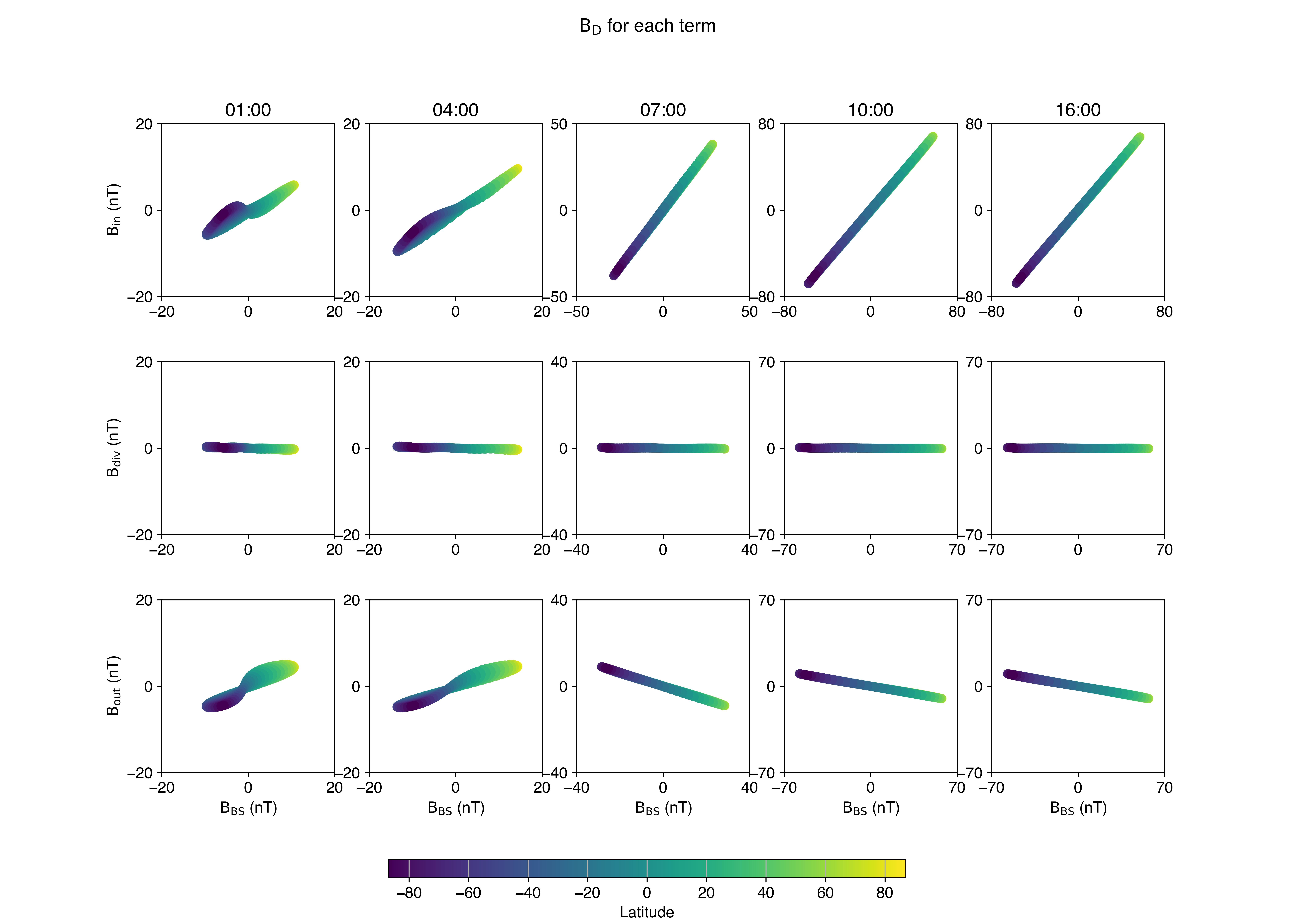}
\caption{$B_D$ from $\innerint$, $\diverr$, and $\outerr$ versus $\bsint$ using OpenGGCM data. Same format as Figure~\ref{fig:OpenGGCM Bn integrals vs Biot}}.
\label{fig:OpenGGCM Bd integrals vs Biot}
\end{figure}

%
%

\clearpage
\bibliography{Bibliography}

\end{document}